\documentclass[lettersize,journal]{IEEEtran}
\usepackage{amsmath}
\usepackage{amsfonts}

\usepackage{graphicx}
\usepackage{array}
\usepackage{float}
\usepackage{stfloats}
\usepackage[
    caption=false,
    font=normalsize,
    labelfont=sf,
    textfont=sf
]{subfig}

\usepackage{algorithm}
\usepackage{algorithmic}

\usepackage{textcomp}
\usepackage{url}
\usepackage{verbatim}

\def\BibTeX{{\rm B\kern-.05em{\sc i\kern-.025em b}\kern-.08em
    T\kern-.1667em\lower.7ex\hbox{E}\kern-.125emX}}
\usepackage{balance}

\newcounter{imagepanel}
\renewcommand{\theimagepanel}{\alph{imagepanel}}

\begin{document}
\title{Informed Sinogram Interpolation for Sparse View Reconstruction}

\author{
  \IEEEauthorblockN{
    Yuejie Liu\IEEEauthorrefmark{1}\IEEEauthorrefmark{2},
    Alessandro Lupoli\IEEEauthorrefmark{1}\IEEEauthorrefmark{4},
    David Uribe Gallo\IEEEauthorrefmark{2},
    Andreas Fischer\IEEEauthorrefmark{2}, and
    Felix Krahmer\IEEEauthorrefmark{1}\IEEEauthorrefmark{3}\IEEEauthorrefmark{4}
    }
  \\
  \IEEEauthorblockA{
    \IEEEauthorrefmark{1}Department of Mathematics, \textit{Technische Universität München}
  }\\
   \IEEEauthorblockA{
    \IEEEauthorrefmark{2}\textit{Waygate Technologies, Baker Hughes Digital Solutions GmbH}\\
  }
  \IEEEauthorblockA{
    \IEEEauthorrefmark{3}Department of Electrical Engineering and Information Technology, \textit{Technische Universität Darmstadt}
  }\\
  \IEEEauthorblockA{
    \IEEEauthorrefmark{4}\textit{Munich Center for Machine Learning} (MCML)
  }\\
  Email: alessandro.lupoli@tum.de
}

\maketitle

\begin{abstract}
Computed tomography (CT) has been widely used in medical examinations and non-destructive testing.
Micro-CT (microfocus X-ray CT system) is an advanced version that can observe the internal structures of small objects. However, secondary radiation can prevent micro-CT from imaging the full structure of the object. Given the same amount of secondary scans as primary scans, simple subtraction solves the problem at the price of doubled acquisition time. To reduce the acquisition time, one aims to limit the secondary scans to as few imaging angles as possible and interpolate to recover the missing data.\\
Different interpolation methods have been explored for limited-angle tomography, such as polynomial or spline interpolation, compressive sensing, deep learning, etc.
In contrast to many of these scenarios, the secondary imaging setup considered in this paper often exhibits simple sinusoidal structures that are not exploited directly by any of the aforementioned approaches.
This paper aims to fill this gap with a continuous mathematical model describing the secondary sinogram only with information from the projection space. Building on this model, we propose an interpolation method that effectively interpolates secondary scans from limited observations.
\end{abstract}


\section{Introduction}
The attenuation of monochromatic X-ray radiation is empirically described by the Beer-Lambert law. 
Given the initial X-ray energy $I_0$, after being attenuated by the object, the output energy is
\begin{equation*}
  I = I_0 e^{-\int_L\mu},\label{eq:attenuation}
\end{equation*}
where $\mu$ is the attenuation coefficient, which depends on
the density and other physical properties of the object \cite{stein2011fourier}.
In parallel-beam geometry, where $L(r,\theta) = \{(x,y)\ | \ x\cos\theta+y\sin\theta=r\}$, the line integral $\int_L\mu$ in the exponent is the Radon transform of $\mu$ \cite{Radon:1917}, as given by

\begin{align*}
    \mathcal{R}\mu(r,\theta) &= \int_{-\infty}^\infty\int_{-\infty}^\infty \mu(x,y) \delta(x\cos\theta+y\sin\theta-r)\text{d}x\text{d}y\\
    &= \int_{-\infty}^\infty \mu(r\cos\theta+s\sin\theta,-s\cos\theta + r\sin\theta) \text{d}s.
\end{align*}
  
$\mathcal{R}\mu(r,\theta)$ is also called a sinogram, as an off-center point is projected as a sinusoid. This sinusoidal structure is preserved by taking exponentials, which is why we refer to the output energy $I$ as the intensity sinogram. The concept of a sinogram, however, can be extended to other imaging geometries, such as cone-beam geometry (see Section \ref{sec:second-sect-intr} for details) due to similar phenomena. The reconstruction of $\mu$ from samples $\mathcal{R}\mu(r_k,\theta_j)$ is a typical inverse problem. In practice, the filtered back projection (FBP) \cite{maier2018medical} and the Feldkamp-Davis-Kress (FDK) algorithm can find $\mu$ for parallel-beam geometry and cone-beam geometry (see \cite{feldkamp1984practical} for circular and \cite{Sourbelle2002} for helical), respectively.

\subsection{Secondary radiation artifacts}%
\label{sec:second-sect-intr}
\noindent In a typical process of X-ray generation, electrons are emitted from a heated filament (negatively charged cathode) and accelerated toward the anode (positively charged target). These electrons are then focused by magnetic coils before striking the metal target, where X-rays are generated \cite{Berger2018}.

Unlike parallel-beam and fan-beam CT, cone-beam CT captures multiple lines in one projection. With the benefit of lower-dose and fast imaging, it is widely employed in medical fields \cite {SCARFE2008707} and non-destructive testing. Micro-CT is a variant of CT that uses a microfocus X-ray tube, typically applied in a cone-beam setup \cite{vasarhelyi2020microcomputed}.  While it follows similar physical principles to standard CT, its highly focused X-rays allow for imaging at microscopic resolution (voxel size 
$\ll 100 \mu m^3$) \cite{BADEA202147} . Similar to standard CT, the reconstruction is obtained by taking hundreds of two-dimensional projections from multiple angles around the sample. 
 Micro-CT systems commonly exhibit inaccuracies caused by off-focal radiation, or secondary radiation, that one encounters in addition to the intended primary radiation \cite{FLAY2020472}\cite{BOONE20127}. In the projection, this can lead to artifacts, and one observes the sum of the desired primary projection and a secondary projection resulting from the off-focal radiation. Using an advanced X-ray tube configuration, it is possible to generate off-focal radiation alone, thus only imaging secondary projections; we will refer to these measurements as secondary scans.
\cite{Kingston:2022} provides a possible explanation and simulations of the off-focal effect for both transmission and reflection targets, where the transmission profile can be modeled by a Lorentzian function. The four possible mechanisms of off-focal X-ray generation are:

\begin{figure*}[t]
    \centering
    \includegraphics[width=.8\linewidth]{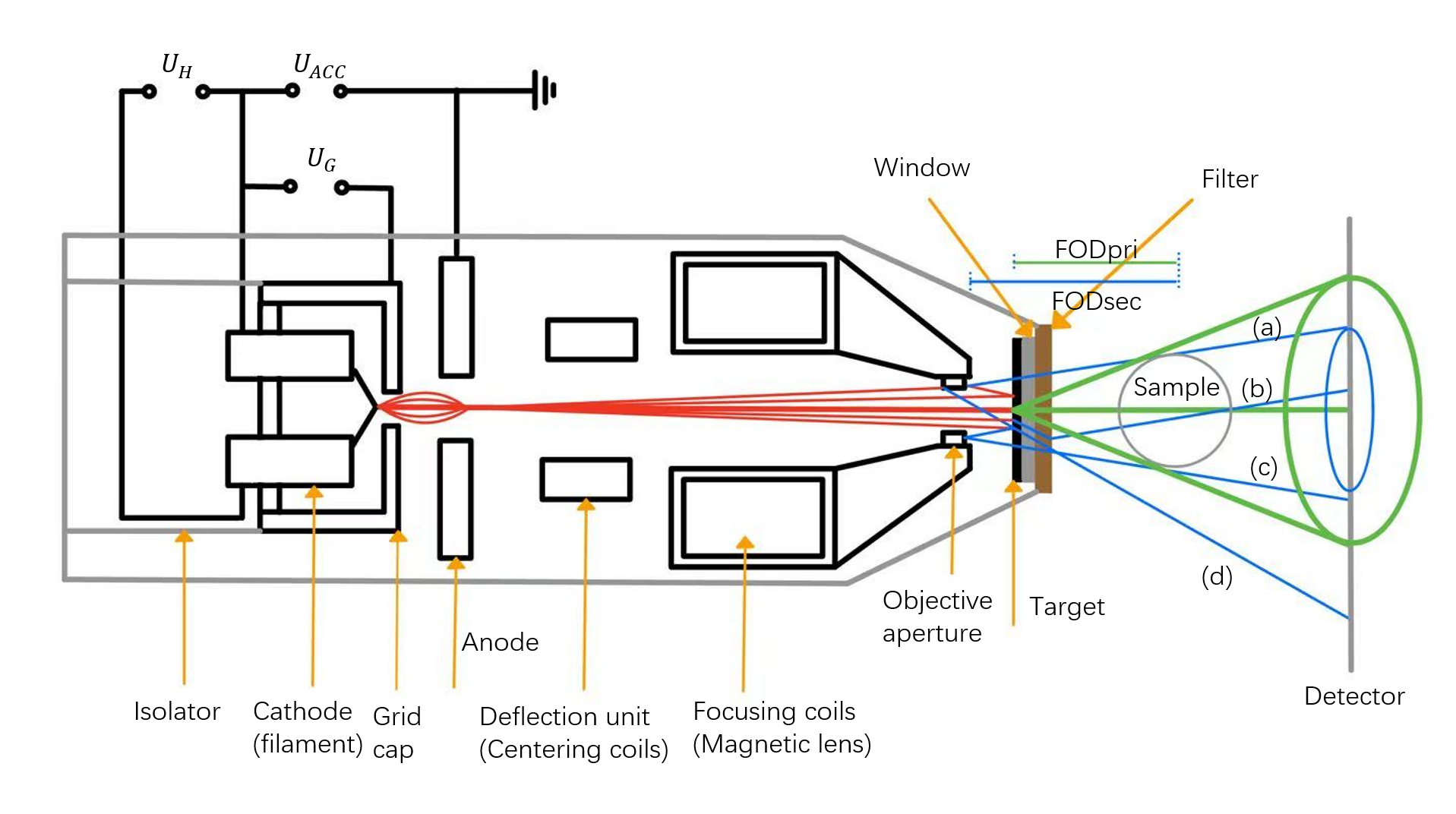}
     \caption{Imaging process of a transmission cone-beam microfocus X-ray tube and four off-focal X-ray mechanisms \cite{websiteXrayHead}\cite{websiteXrayTubeBH}\cite{websiteXrayTubeWorX}\cite{Kingston:2022}. $U_H$: filament heating voltage; $U_{ACC}$: acceleration voltage; $U_{G}$: grid control voltage. FOD: focal object distance. Red lines: electron beam; Green lines: primary X-rays; Blue lines: secondary X-rays.}
  \label{tube}
\end{figure*}

\begin{enumerate}
    \item[(a)] Part of the electrons are backscattered from the target and interact with internal components within the X-ray tube, leading to the generation of additional bremsstrahlung radiation (see also \cite{FLAY2020472}\cite{BOONE20127});
    \item[(b)] forward-propagating focal X-rays may be scattered within beam-filtering materials—such as the aluminum filter—positioned near the source;
    \item[(c)] Some focal X-rays are redirected backward within the gun assembly, where they can be deflected toward the detector;
    \item[(d)] stray electrons intercepted by the electron-beam focusing optics or apertures can produce bremsstrahlung X-rays.
\end{enumerate}

\begin{figure*}[t]
    \centering
    \setcounter{imagepanel}{0}

    \begin{minipage}[t]{0.31\textwidth}
        \centering

        \refstepcounter{imagepanel}
        \label{Primary Scan}

        \includegraphics[width=\linewidth]{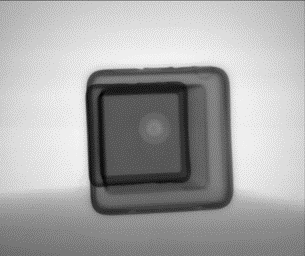}

        \smallskip
        {\small (\theimagepanel) Primary scan at $0^\circ$}
    \end{minipage}
    \hfill
    \begin{minipage}[t]{0.31\textwidth}
        \centering

        \refstepcounter{imagepanel}
        \label{Secondary Scan}

        \includegraphics[width=\linewidth]{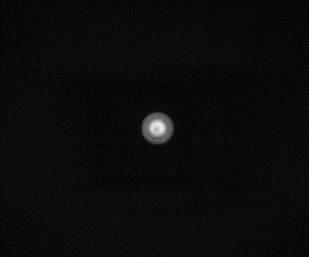}

        \smallskip
        {\small (\theimagepanel) Secondary scan at $0^\circ$}
    \end{minipage}
    \hfill
    \begin{minipage}[t]{0.31\textwidth}
        \centering

        \refstepcounter{imagepanel}
        \label{ReconCTnocorr}

        \includegraphics[
            width=\linewidth,
            height=4.675cm
        ]{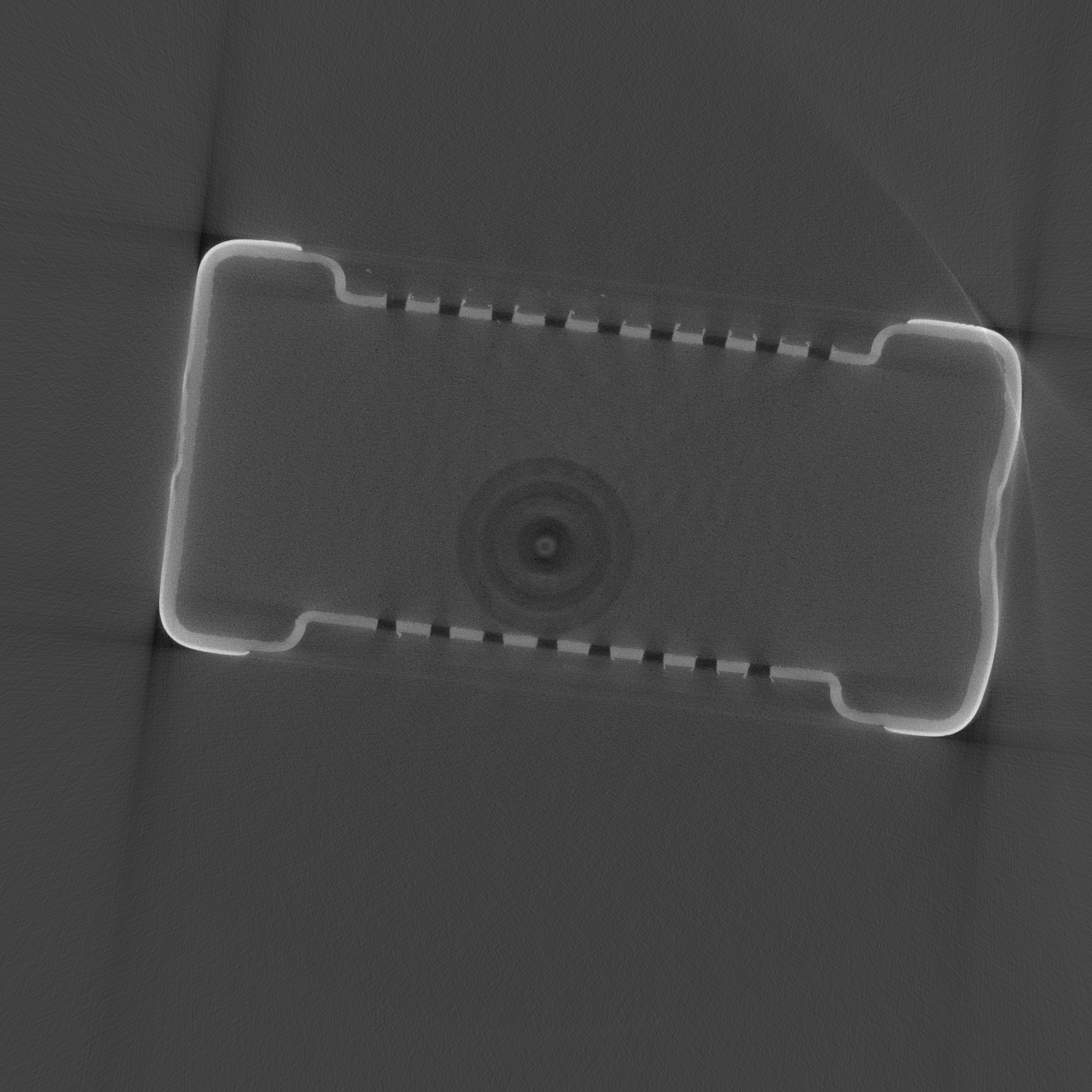}

        \smallskip
        {\small (\theimagepanel) Reconstructed CT (slice $1251$ of $2502$)
        on the $z$-axis}
    \end{minipage}

    \caption{Primary and secondary scans of a coil and the reconstructed CT
    slice based solely on primary scans.}
    \label{fig:coil-scans}
\end{figure*}

Even though the imaging principle has not been fully understood, our observations from the scans suggest that this localized artifact results from larger focal object distance than the primary radiation, a condition corresponding to cases (a), (c), and (d).
When a full sequence of secondary scans is available, a clean primary projection without the artifact can be obtained by subtracting the secondary scan from the corresponding primary scan. However, obtaining these secondary scans
doubles the acquisition time. In this paper, we will demonstrate that it is often sufficient to acquire secondary scans for just a limited number of angles, and then interpolate the missing angles to obtain a full sequence of secondary scans that can then be used to enhance the corrupted primary scans. 

\begin{figure}\label{secondary_of_the_scanned_object}
    \centering
    \includegraphics[width=1\linewidth]{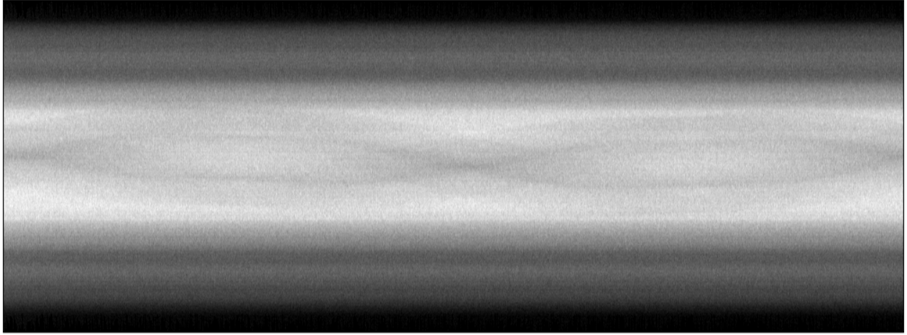}
    \centering
  \caption{Middle level $h = \lfloor \frac{H+1}{2} \rfloor$ secondary intensity sinogram of a coil (Figure~\ref{fig:coil-scans}\ref{Secondary Scan}) after cropping the donut with a bounding box with height $H+1$}
  \label{Secondary_Sinogram}
\end{figure}

For transmission microfocus X-ray tubes, which are the motivating application of this paper, these off-focal source effects are isotropic \cite{Kingston:2022}. That is why the artifact caused by off-focal radiation often has a strong radially symmetric donut-shaped component, with blurred, highly attenuating parts of the sample visible within it (see Figure \ref{Secondary Scan}). Similar artifacts in scans and reconstructed CT slices have been investigated (see Figures 7 and 8 in \cite{BOONE20127}), even though the specific artifact varies due to different X-ray tube designs and materials used. In addition, the intensity sinogram representation of the secondary scan (see Figure \ref{Secondary_Sinogram}) shows that these off-focal X-rays together resemble the behavior of parallel-beam X-rays.
\subsection{Image processing ingredients}
\label{Sec:ImageProcessingIngredients}
In the secondary intensity sinogram (Figure \ref{Secondary_Sinogram}), besides the sinusoidal structure, there is an additional constant background, representing the original X-ray energy not influenced by the sample. An approximation of this background can be obtained via the so-called (secondary) background-only image, which is obtained by placing no sample
 in front of the X-rays. We denote the secondary background-only image by \(I_{0, sec}\). Analogously, we will denote the primary background-only image by \(I_{0,pri}\). 
 
 Background-only images are commonly used to reduce fixed-pattern noise before applying a reconstruction algorithm, such as FBP or FDK. The conventional normalization technique employed is the flat field correction (FFC) \cite{VanNieuwenhove:15}, defined as
\begin{equation}\label{main_correction}
  I_{corrected,FFC} = \frac{I-I_{offset}}{I_0-I_{offset}},
\end{equation}
where $I_{offset}$ is the offset image captured by the detector without X-ray illumination. Given that absorption of the object reduces the X-ray energy, one typically has $I\leq I_0$ for small enough noise levels, 
and the pixel values of $I_{corrected, FFC}$ are between $[0,1]$. To adjust this normalized image to the original scale range, \cite{ParkSharp:2016} corrects the image by multiplying the right-hand side of \eqref{main_correction} with the mean value of the denominator.
For cases where the data distribution is skewed or outliers exist, median weighting is better than the mean value. Consequently, we will use a median-based correction can denote the corrected image as $I_{corrected}$ 
\begin{equation}
  I_{corrected,median} = \frac{I-I_{offset}}{I_0-I_{offset}}median(I_0-I_{offset}).
\end{equation}

\begin{figure}[H]
    \centering

    \begin{minipage}[t]{0.48\linewidth}
        \centering
        \includegraphics[width=\linewidth]{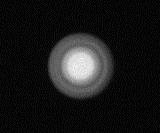}

        \smallskip
        {\small (a) Normalized secondary background-only image}
    \end{minipage}
    \hfill
    \begin{minipage}[t]{0.48\linewidth}
        \centering
        \includegraphics[width=\linewidth,height =3.525cm]{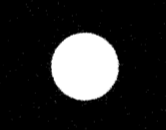}

        \smallskip
        {\small (b) Normalized secondary background-only image after binary thresholding}
    \end{minipage}

    \caption{The process of mask creation, with a closer look at the central area.}
    \label{masks}
\end{figure}

The FDK algorithm is applied to $I_{corrected}$ to reconstruct the attenuation coefficient of the sample.
Furthermore, one can exploit that the main part of the artifact occupies only a small portion of the image (see Figure \ref{fig:coil-scans}\ref{Secondary Scan}); therefore, we can create a mask to eliminate the noise around the donut-shaped object,
and a corresponding bounding box that cuts out the useful information.
The mask is created mainly by normalizing the gray values of the secondary background-only image to $[0,255]$ and applying a binary threshold (see Figure \ref{masks}).

\subsection{Modeling fundamentals}
The main challenge with the off-focal radiation is that there is no apparent clear and unique focal spot and its specifics can depend on multiple hardware parameters, see Section~\ref{sec:second-sect-intr} above. 
Consequently, to the best of our knowledge, there is no model of the secondary imaging process justified by physical principles. 
Empirically, however, the structure of sinusoids of the secondary sinogram resembles the one produced by the Radon transform under parallel-beam geometry. To reflect this, the aim of this paper is to build a general continuous mathematical model for the secondary intensity sinogram 
which can then be exploited to interpolate missing angles. The precise setup of the model is driven by the following considerations. 

In general, a grayscale image $F$ is a map from a 
domain $\Omega \subset \mathbb{R}^{2}$ to $S = [a,b]\subset\mathbb{R}$.
To store images in computers, $F$ is often discretised with $\Omega^{'}=\{(k,j)\ |\ k\in \{0, \dots, K\}, j \in \{0, \dots, J\}\} \subset \mathbb{N}^2$ and 
$S^{'} = \{0, 1, 2, \dots, 2^{N}-1\}$, which is essentially a matrix. For example, in our dataset of 16-bit TIFF images, the maximum value that can be stored in an image is $2^{16}-1=65535$. 

A dataset of X-ray scans has three dimensions: two representing the dimensions of the projections and one representing the angles. The primary and secondary scans represent the detected values at the corresponding locations at a detector array, and are indexed accordingly by $[0, H^{'}_{cont}]\times[-R^{'}, R^{'}]$, where the detector height is denoted by $H^{'}_{cont}$ and the width by $2R^{'}$, both in millimeters.
Note that the detector array actually generates discrete digital images, so the continuous domain of the primary scans is discretised as $\{0,1,..., H^{'}_{disc}\}\times \{0,1,..., I\}$, where $H^{'}_{disc}\in \mathbb{N}_+,\ (H^{'}_{disc}+1)\times pixel\_size=H^{'}_{cont}, \ and\  (I+1)\times pixel\_size=2R^{'}$. The discrete secondary scans follow similarly.

A discrete primary scan is acquired for $J+1$ different rotation angles, which we denote  as $$\Theta = \{\theta_0, \theta_1,\dots, \theta_J\ |\ \theta_0<\theta_1<\dots<\theta_J,\ \theta_0 = 0, \theta_J = 2\pi\}.$$ To remove the artifacts resulting from off-focal radiation, one aims to subtract from each of these primary scans the corresponding secondary scan. The goal of this paper is to investigate the scenario where only parts of these secondary scans are available, and the missing ones need to be interpolated. More precisely, we assume that we have access to the subsampled set of angles  
\begin{equation*}
\Theta_{\mathrm{sample}}
=
\left\{
\theta_{j_0},\theta_{j_1},\ldots,\theta_{j_N}
\;\middle|\;
\begin{aligned}
\theta_{j_0} &< \theta_{j_1} < \cdots < \theta_{j_N},\\
\theta_{j_0} &= \theta_0 = 0,\\
\theta_{j_N} &= \theta_J = 2\pi,\qquad N \leq J
\end{aligned}
\right\}.
\end{equation*}

Our general strategy is to interpolate the secondary intensity sinograms for different rows of the scans (i.e., different level parameters $h$) sequentially, but incorporate the smoothness in $h$ by using the reconstructions already completed as initializations for the following ones. To find a good starting point, we compute an interpolation of secondary absorption sinograms (see Section \ref{sec:initmod}) for some candidate rows and evaluate their quality.

Let $F$ be the secondary intensity sinogram for a candidate row to be interpolated. 
Since the artifact information is primarily concentrated in the central portion of the secondary scans, we will crop these scans to the bounding box dimensions $H_{cont}\times(2R)$ and focus only on the cropped discrete secondary intensity sinograms. The information we have available consists of the samples 
\[
F(r_k,\theta_{j_n}),\ k = 0,1,...,K,\ n = 0,...,N,
\] where $r_k = \frac{k}{K},\ k=0,\dots, K$ represents the normalized indices of the cropped rows.
The vertical cropping dimension is denoted by $H_{cont}=(H+1) \times pixel\_size$ and corresponds to the height of the cropped secondary scans. Hence, we aim to recover a total number of $H+1$ secondary intensity sinograms.

For each of these sinograms, we aim to complete the array of gray values $F(r_k,
\theta_{j}),\ k=0,\dots, K,\ j = 0,\dots, J$, i.e., the values for all angles in $\Theta$ from the values for angles in $\Theta_{sample}$; here $r_k$ is not subsampled.
In our case, the detector elements have the same size and the rotation angles are equidistantly sampled (both $\Theta$ and $\Theta_{sample}$).
Since the rotation period is always $2\pi$, $\Theta = \{\theta_j\ |\ \theta_j= \frac{2j\pi}{J},j = 0,\dots,J\}$, $\Theta_{sample} = \{\theta_{j_n}\ |\ \theta_{j_n}= \frac{2n\pi}{N},n = 0,\dots,N\}$

\begin{figure}
  \includegraphics[scale = .24]{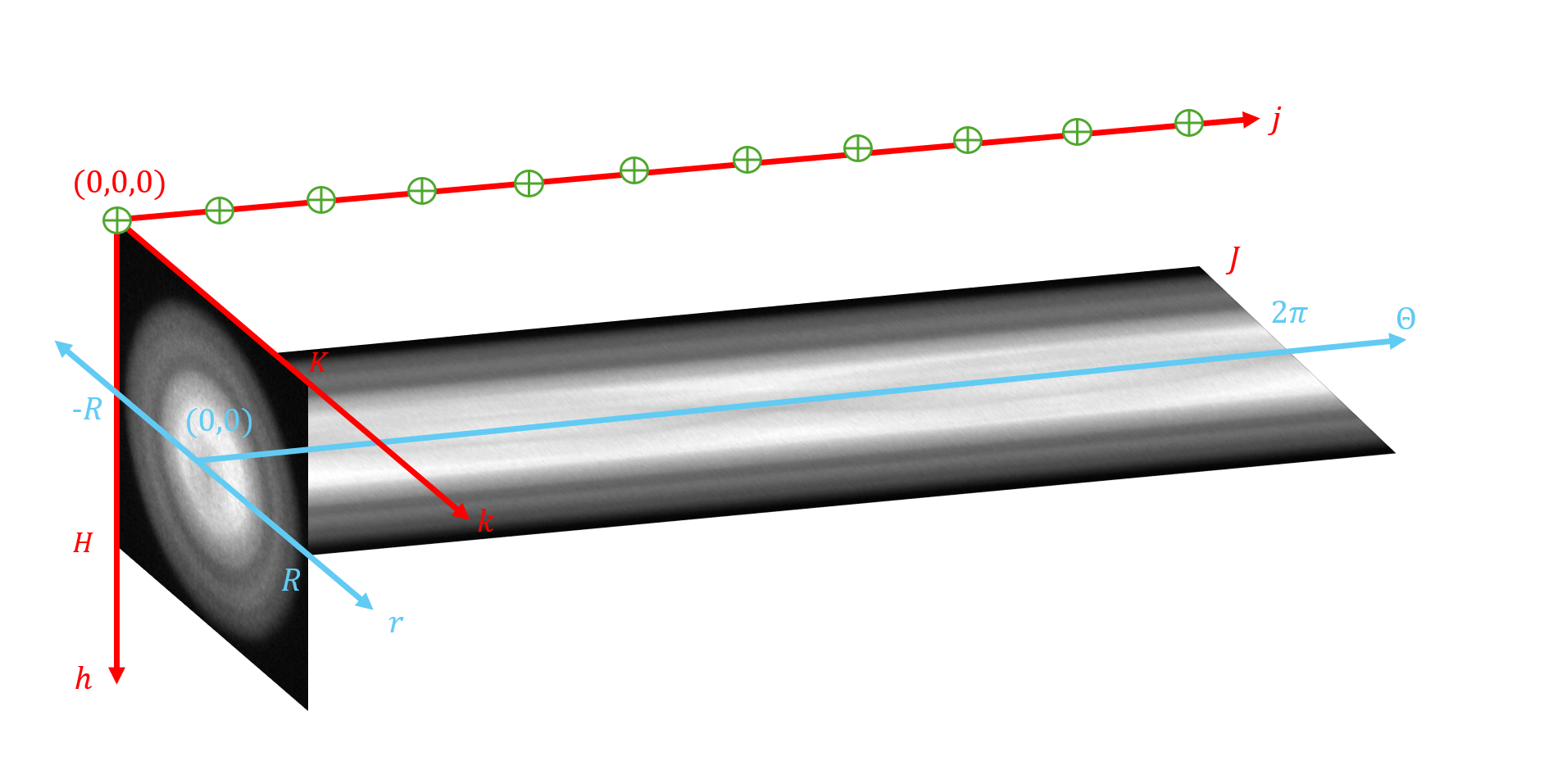}
  \centering
  \caption{The coordinate systems of the cropped secondary scan ($\theta_j= 0$) and the middle secondary intensity sinogram (level $h = \lfloor \frac{H+1}{2}\rfloor$)}
  \label{SinoCoord}
\end{figure}

 Figure \ref{SinoCoord} visualizes the coordinate systems of secondary scans after masking and cropping by the bounding box:
\begin{itemize}
  \item blue coordinate: the continuous coordinate system of the cropped secondary intensity sinogram.
  \item red coordinate: the discrete coordinate system of the cropped secondary scan and intensity sinogram, the same as in a computer. $(H+1) \times pixel\_size = H_{cont},\ (K+1) \times pixel\_size =2R\ (unit: mm)$, $J\times angular\_discretization\_step = 2\pi$. 
  \item green crosses with circles: an example of the subsampled set of angles ($J = 1000,\ N = 10$).
\end{itemize}

\section{Previous Work}
In the context of parallel-beam imaging, a number of methods for sinogram interpolation have been explored. In particular,  one can consider each pixel in the detector array separately and apply one-dimensional interpolation methods, as they have been intensively studied in approximation theory. 
More concretely, in \cite{Talha2017InterpolationBE}, four standard interpolation methods are considered -- nearest neighbor, linear, spline, and Hermite interpolation --
and their application to parallel-beam sinogram interpolation is investigated. 
In \cite{BenammarAllagDraiYahiBoutkedjirt:2019}, the pixel-wise interpolation is performed via the robust smoothing approach proposed in \cite{GARCIA20101167} in combination with an iterative refinement strategy in the angular domain.

However, the structure of the sinogram is not explicitly used in pixel-wise interpolation. Only a few empirical methods have been proposed that directly utilize the sinogram structure.
\cite{Widodo} assumes the values on the same sine wave are close, and spline interpolation is employed 
along the empirical sine waves after grouping similar gray values for given angles.
\cite{MarttiKalke} proposed the method SINT (Sinogram Interpolation Technique), which is based on an appropriately weighted sum of certain candidate sine waves specifically designed for modeling parallel-beam CT sinograms.  \cite{Zeng:2019} introduces a displacement function interpolation for the sinogram, which has the advantage that it generalizes to other imaging geometries.
Directional interpolation is also a popular method, see \cite{MarttiKalke} for a more detailed introduction, both for the general case and the sinogram case. 
For example, \cite{Bertram2009DirectionalVI} adopts directional interpolation for the sinogram by exploring the 3D information and is also not limited to imaging geometries. 

When the angles are subsampled at random and the Fourier transform (in the angular variable) of the sinogram data is sparse, as it is empirically observed for certain CT scenarios, one can also apply compressive sensing to recover missing sinogram data  \cite{6923935}. 
This method, however, is not applicable to our problem: most importantly, the secondary sinograms do not exhibit a similar Fourier sparsity structure as observed for the scenarios considered in \cite{6923935} (or if they do, this structure is almost entirely covered by noise); in addition, our projection data is equidistantly sampled rather than at random.

When the sinogram information for a larger fraction of the angles is available, one can also employ variational methods. \cite{Kstler2006AdaptiveVS} formulates an energy function that
uses a gap mask to ensure the observed sinogram and ideal sinogram are equivalent at known data positions,
with a linear regularizer to fill the missing information. The corresponding PDE with Neumann boundary conditions
is discretized and solved using various methods. The experiments are performed on simulated parallel-beam data with 60\% missing angles. 
A similar approach is applied to simulated and real cone-beam data with 50\% missing angles in \cite{Karimi2021InterpolationOC} 
employing a regularization driven by self-similarity and smoothness of sinograms. 

Recently, deep learning methods have gained a lot of attention due to their ability 
to adapt to all kinds of downstream tasks with typical backbones like convolutional neural networks (CNN) and 
generative adversarial networks (GAN). Although without theoretical guarantees in most scenarios, deep learning
has shown its potential to solve sinogram interpolation for real-world data. For example, 
\cite{10.1117/12.2512979} combines U-Net and residual learning to improve linear interpolation for micro-CT projections;
Bellens et al. first develop a sinogram representation called spinogram, then feed it into a CNN with spinogram metadata, and 
validate on simulated cone-beam CT images\cite{Bellens2023AML}. In their consecutive work \cite{BELLENS2024449},
the validation is extended to real X-ray CT projections. 

Despite the amazing results deep learning has achieved, it has three problems. First, the model relies on the training dataset and may not be applicable to all situations.
Second, deep learning typically requires a large dataset, while there is no such dataset for the secondary projections.
Finally, although deep learning can fit a function that portrays the structure of sinograms, it is computationally expensive.

Another area of related work is sinogram inpainting, where the goal is to restore relatively large, irregularly shaped missing regions within the sinogram domain. This kind of problem often occurs due to severe physical limitations such as metal artifacts (see \cite{Peng2017,6153697,Chen2012CTMA,5652149,7485845})
or data truncation (see
\cite{Li2019ASI,app12126268,lehan_2024,s19183941,Tovey_2019,Zhao2018UnsupervisedLS}). The concept of inpainting is closely related to interpolation (see \cite{LI20142862,Wagner2023GeometricCE,11084648,Jiaze2024FCDMAP,e2026trainingfreeinferencehighresolutionsinogram,doi:10.2352/EI.2025.37.12.HPCI-192}), so powerful inpainting methods could also be leveraged to tackle interpolation challenges. However, most of these methods are based on deep learning techniques, which makes them difficult to adapt to secondary sinograms due to the lack of sufficient amounts of training data.

\section{Algorithmic ingredients}
\subsection{Background separation and initial model}
\label{sec:initmod}
Our main goal is to describe the sinusoid structure in a continuous mathematical model for interpolation. To begin, we will subtract the secondary background-only image $I_{0,sec}$ to remove the constant background in each secondary intensity sinogram $I_{sec}$ to obtain the secondary absorption sinogram $f(r,\theta):=I_{0,sec}-I_{sec}$.

\begin{figure}[htbp]
  \centering
  \includegraphics[scale = .26]{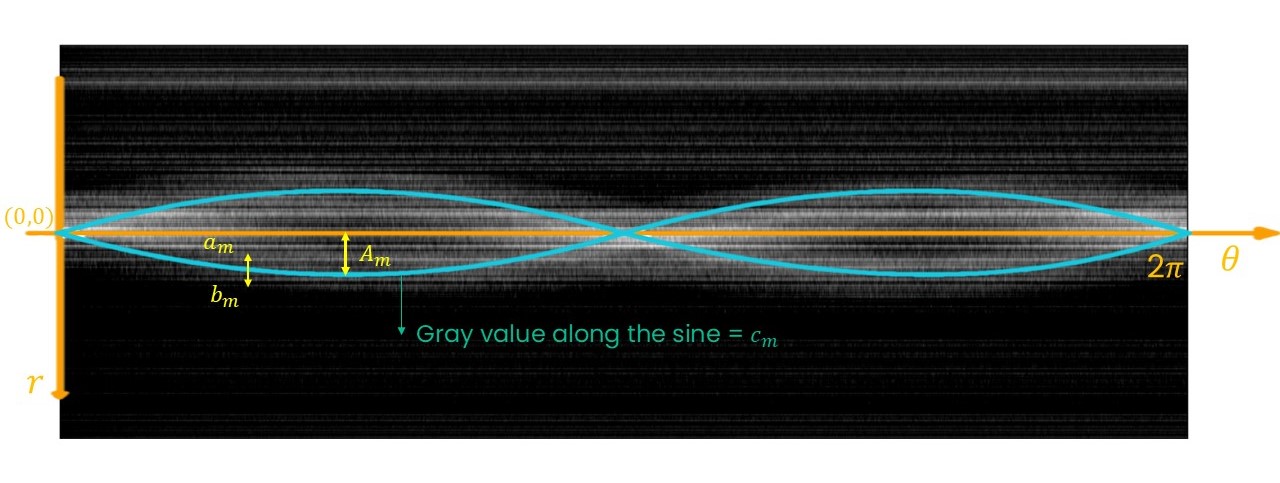}
  \caption{Illustration of the initial continuous model of a secondary absorption sinogram}
  \label{MinusSecSino}
\end{figure}

Since $1-e^{-x}$ is a monotonically increasing function of $x$, the level sets of $f(r, \theta)$ qualitatively (up to sign) resemble those of the secondary sinogram $g(r,\theta)$, which is defined in analogy to the parallel-beam setup discussed above, implicitly via
$I_{sec} = I_{0,sec}e^{-g(r,\theta)}.$ 
The resemblance to the parallel-beam setup that modifies the term sinogram is exemplified by Figure \ref{MinusSecSino}: the secondary (absorption) sinogram indeed resembles a blurred parallel-beam sinogram. In particular, when the object has highly attenuating parts, one observes sinusoids as they arise in the parallel-beam geometry when point masses are present. More precisely, in parallel-beam geometry, if tiny objects are abstracted as masses  $\mu(x,y) = \delta_{\{(x_0,y_0)\}}(x,y)$,
they generate sinusoids with a constant gray scale in the sinogram.
Similarly, there are only two obvious sine waves in the secondary absorption sinogram in Figure \ref{MinusSecSino}, reflecting that the sample contains two highly absorbing parts.

Thus, in analogy to the parallel-beam setup, if the sample has tiny well-separated regions of high absorption, one expects the non-blurred sinogram to display a finite union of sine curves with different amplitudes, phases, offsets, and gray scales, each of which can be described via an equation of the form $r = A\sin(\theta+\phi)+\beta$. Consequently, we model the non-blurred secondary sinogram as follows:
\begin{equation}
  p(r,\theta) = \sum_{m=1}^{M}c_m\delta(r-A_m\sin(\theta+\phi_m)-\beta).
\end{equation}
Here we use the following notation and normalization.
\begin{itemize}
  \item $r\in[-R,R]$ and $\theta\in[0,2\pi]$
  \item $A_m$: amplitudes of the sine waves
  \item $\phi_m$: phases of the sine waves
  \item $\delta$: Dirac delta measure, centered at $0$
  \item $c_m$: the gray values along a single sine wave determined by the corresponding amplitude and phase
  \item $\beta$: the offset of the imaging center
\end{itemize}
To include the blur into the model, we incorporate a convolution with a kernel $K_{\sigma}$.
Thus, we have
\begin{align}
    f(r,\theta) &= \sum_{m=1}^{M}c_m \mathcal{K}_\sigma(r-A_m\sin(\theta+\phi_m)-\beta_m),\label{eq:fc}\\
\end{align}
where $\mathcal{K}_\sigma(r) = \mathcal{K}(\frac{r}{\sigma})$ is a convolution kernel, whose scaling factor $\sigma$ is assumed to be known.

If we write $s_m(\theta) = A_m\sin(\theta+\phi_m)+\beta$, then the blurred sinogram model takes the form
\begin{equation}
  \begin{aligned}
      f_\theta(r) &= \sum_{m = 1}^{M}c_{m}\mathcal{K}_{\sigma}(r-s_m(\theta)) = \mathcal{K}_{\sigma}\mathop{*}p_\theta(r),\\
  p_\theta(r) &= \sum_{m = 1}^{M}c_{m}\delta_{s_m(\theta)}(r).
  \end{aligned}
  \label{PulseModel}
\end{equation}
A strategy to solve this system is to apply the Fourier transform, which translates the convolution into a multiplication. If the kernel $\mathcal{K}_\sigma$ is non-vanishing in Fourier space, computing the inverse Fourier transform of 
$\hat{p}_\theta(\xi) =\frac{\hat{f}_\theta(\xi)}{\hat{\mathcal{K}}_{\sigma}(\xi)}$ will lead to the solution.

However, similarly to the parallel beam scenario, one has access to the values of $p$ only on some finite grid $\{r_0, \dots, r_K\}\times \{\theta_{j_0}, \dots, \theta_{j_N}\}$, so the observations are given as
\begin{equation}
 f_{\theta_{j_n}}(r_k) = \sum_{m = 1}^{M}c_{m}\mathcal{K}_{\sigma}(r_k-s_m(\theta_{j_n})) = \mathcal{K}_{\sigma}\mathop{*}p_{\theta_{j_n}}(r_k).
\end{equation}
For each $\theta_{j_n}$, finding the $c_m$'s and the $s_m(\theta_j)$'s from these observations is a  \textbf{Prony type problem}, which is well studied in the mathematical literature (see \cite{Sauer2018-ky}\cite{Plonka2023}) and for which also a number of efficient algorithms have been designed such as MUSIC (MUtiple SIgnal Classification) \cite{LIAO201633},
ESPRIT(Estimation of Signal Parameters via Rotational Invariance Techniques) \cite{fannjiang2016compressivespectralestimationsinglesnapshot}, and convex optimization \cite{BENDORY2016511}. However, for this problem to be even well-posed, one needs the peaks $s_m(\theta_j)$ to be well-separated, which cannot be guaranteed for all sampled angles and all overlayed sine curves - and one numerically observes that even a few violations of these assumptions prevent the success of the overall recovery.

\subsection{Towards a more practical model}
In the previous section, the highly absorbing parts are abstracted as mass points, omitting volume information. To include the volume information in the model, the sinusoids in a non-blurred secondary absorption sinogram should have widths, which inspires us to convolve each sinusoid with a symmetric characteristic function $\chi_[-\epsilon_m,\epsilon_m](r)$:
\begin{equation}
  \bar{p}_{\theta}(r)= \sum_{m=1}^{M}c_m\chi_[-\epsilon_m,\epsilon_m](r-A_m\sin(\theta+\phi_m)-\beta).
\end{equation}
As a result, we propose a continuous model of the secondary absorption sinograms by replacing the Dirac delta measure with characteristic functions $\chi_{[a,b)}(x)$. The shift $\beta$ is removed because the interval $[a_m,b_m)$ can include the offset information of the imaging center. These asymmetric intervals also allow more complexity of the model: the width representation of each sinusoid need not be centered around $s_m(\theta_j)$.

\begin{equation}\label{eq:final}
\begin{aligned}
  f(r,\theta) &= \mathcal{K}_\sigma\mathop{*}\bar{p}_{\theta}(r) \\
  &=\mathcal{K}_\sigma\mathop{*}[\sum_{m = 1}^{M}c_{m}\chi_{[a_m,b_m)}(r-A_m\sin(\theta+\phi_m))].
\end{aligned} 
\end{equation}

To reflect the parameter range limitations observed in a secondary absorption sinogram, a few constraints need to be added. To incorporate in the sinogram model that  $\beta$ is small, but does not need to vanish, we impose that $0\in[a_m, b_m]$, so we obtain the constraints 
\begin{equation}
  \begin{aligned}
    0\leq &A_m\leq R,\\
    0\leq &\phi_m \leq 2\pi,\\
    -R \leq &a_m\leq 0,\\
    0 \leq &b_m \leq R, \\
    0\leq &c_m \leq \max_{r,\theta}{f(r,\theta)}.
  \end{aligned}
  \label{boxConstraints}
\end{equation}
Considering the translation of the coordinate systems, the corresponding discrete model is 
\begin{equation}
\begin{aligned}  
    f(k,\theta_j) &= \mathcal{K}_\sigma\mathop{*}[\sum_{m = 1}^{M}c_{m}\chi_{[a_m,b_m)}(k-A_m\sin(\theta_j+\phi_m))],\\
    &0\leq A_m\leq \left\lfloor \frac{K+1}{2} \right\rfloor,\\
    &0\leq \phi_m \leq 2\pi,\\
    &0\leq a_m\leq \left\lfloor \frac{K+1}{2} \right\rfloor,\\
     &\left\lfloor \frac{K+1}{2} \right\rfloor \leq b_m \leq \lfloor K+1 \rfloor, \\
    &0\leq c_m \leq \max_{r,\theta}{f(r,\theta)},
  \label{eq:finaldiscrete}
  \end{aligned}
\end{equation}
where one can use $ \max_{r,\theta}{f(r,\theta)} = \max_{k,n}{f(k,\theta_{j_n})}+C$ in practice.


Consequently, the square loss between the secondary absorption sinogram data  $\tilde{\mathbf{f}} = (\tilde{f}(k,\theta_{j_n}))$ to be interpolated and the model predictions  $\mathbf{f} = (f(k,\theta_{j_n}))$ for a given parameter vector $\mathbf{x} = (\mathbf{A},\mathbf{\phi},\mathbf{a},\mathbf{b},\mathbf{c}),\mathbf{A} = \{A_m\}_{m=1}^M, \mathbf{\phi} = \{\phi_m\}_{m=1}^M, \mathbf{a}=\{a_m\}_{m=1}^M,\mathbf{b}=\{b_m\}_{m=1}^M,\mathbf{c}=\{c_m\}_{m=1}^M$,  can be expressed in terms of the Frobenius norm when viewing the array of samples as a matrix:

\begin{equation}
\begin{aligned}
&\mathcal{L}(\mathbf{x})
= \left\lVert
\tilde{\mathbf{f}}-\mathbf{f}
\right\rVert_F^2
= \sum_{k=0}^{K}\sum_{n=0}^{N}
\Bigg[
\tilde{f}(k,\theta_{j_n})
\\
&
- \mathcal{K}_{\sigma} *
\Bigg(
\sum_{m=1}^{M}
c_m\,
\chi_{[a_m,b_m)}
\Bigl(
k-A_m\sin(\theta_{j_n}+\phi_m)
\Bigr)
\Bigg)
\Bigg]^2.
\end{aligned}
\label{eq:lossfinal}
\end{equation}

The goal is now to choose the parameter vector in such way that this loss is small.
In this paper, we will focus on the Cauchy kernel $\mathcal{K}_\sigma=\frac{1}{1+(\frac{r}{\sigma})^2}$, a simplified Lorentzian function, with which \cite{Kingston:2022} demonstrates a good simulation of the off-focal effect, but we expect that our approach is not specific to this choice. As the Cauchy kernel $\mathcal{K}_\sigma$ is smooth, it is compatible with gradient-based constrained non-linear optimization methods such as L-BFGS-B \cite{LBFGSB1995}.


There are a few benefits that come with the resulting model:
\begin{enumerate}
  \item These box constraints are not rigid and can be adapted for different sinograms; therefore, 
  even incomplete sinograms with $A_m>R$ or large $\beta$  can be included in the model.
  \item The parameters $\beta$ are hidden in $a_m$ and $b_m$, thus only $5M$ parameters are needed for optimization, which has an accelerating effect compared to applying adjusted column-wise \cite{MarttiKalke} or row-wise \cite{6923935} recovery methods to secondary absorption sinograms.
  \item 
   Compared to the initial model in Section \ref{sec:initmod} that requires solving for $s_m(\theta_j)$ and $A_m$, $\phi_m$ separately, the practical model deals with the two variables $\theta$ and $r$
      together while incorporating the information of $A_m$ and $\phi_m$, thus only a single optimization routine is necessary for one sinogram.
  \item The Cauchy kernel has a primitive function explicitly given as $\sigma\arctan(\frac{r}{\sigma})$, which can be exploited to accelerate the optimization.
   In contrast, the use of other kernels, such as the Gaussian kernel, would entail the need for numerical integration on the interval $[a_m,b_m)$. 
\end{enumerate}

\subsection{Constant background}
However, the na\"ive approach of 
 reconstructing the sinusoids, and inverting the background separation procedure discussed in Section~\ref{sec:initmod}  by subtracting the interpolated secondary absorption sinogram from the corresponding row values of the secondary background-only image is not ideal, as localized errors in the secondary background-only image will lead to line artifacts in the reconstruction.


To address this issue,  we will only apply the aforementioned model to identify the sinograms for the candidate rows, and after selecting one of them as a starting point, we reintroduce the background information into the model, thus considering the secondary intensity sinogram 
\begin{equation}
\begin{aligned}
    F(k,\theta_j) &= d_k\\
    &-\mathcal{K}_\sigma\mathop{*}\left[\sum_{m = 1}^{M}c_{m}\chi_{[a_m,b_m)}(k-A_m\sin(\theta_j+\phi_m))\right],
\end{aligned}
  \label{eq:discreteBG}
\end{equation}
where $0\leq d_k\leq \max_{h,k}{I_{0,sec}(h,k)}$ and for the other parameters, the exact same constraints as before are applied.
The corresponding loss function $\mathcal{L}(\mathbf{x},\mathbf{d})$ simply replaces the model $\mathbf{f}$ and sinogram data $\tilde{\mathbf{f}}$ in (\ref{eq:lossBayesian}) with $\mathbf{F}$ and $ \tilde{\mathbf{F}}$. For this refined model, we apply the L-BFGS-B algorithm\cite{LBFGSB1995}. The natural initialization for $d_k$ is the corresponding gray value of the secondary background-only image. For the other variables, the initialization for most sinogram levels is provided by previously computed neighboring levels, only for a few starting levels, additional parameter tuning is required, see Section  \ref{Sec: Algorithm} for more details. 

\subsection{Incorporating 3D information}
To interpolate all the levels of the secondary scan dataset, we need to recover $H+1$ sinograms. As manually initializing all the parameters for each sinogram is inefficient, it is highly desirable to utilize the 3D information of the dataset. To exploit the observation that the sinusoids change continuously in the $h$ direction,
we add some regularization terms to the loss function to prevent abrupt changes of the parameters. Because the constant background values $d_k$'s exhibit sharper transitions at the top and bottom of the donut ($h$ close to $0$ or $H+1$), we do not include them in the regularization. This method is inspired by the idea of the Kalman filter --
the current system stage is estimated using the current observation combined with the estimate obtained at the previous time instance. 
In our case, the current system state is the underlying model at level $h$, the previous estimate is the model at level $h-1$,
and the current observation is the partially observed sinogram $\tilde{\mathbf{F}}^h$ at level $h$. Initializing the current model $\mathbf{F}^h$ with parameters
from level $h-1$, we have 
\begin{equation}
  \begin{aligned}
    \mathcal{L}(\mathbf{x}^{h},\mathbf{d}^h) &= \Vert \tilde{\mathbf{F}}^h - \mathbf{F}^h\Vert_F^2 + \lambda_1\Vert \mathbf{A}^{h}-\mathbf{A}^{h-1} \Vert_2^2 \\
    &+ \lambda_2 \Vert \mathbf{\phi}^{h}-\mathbf{\phi}^{h-1} \Vert_2^2
    + \lambda_3\Vert \mathbf{a}^{h}-\mathbf{a}^{h-1} \Vert_2^2 \\
    & + \lambda_4 \Vert \mathbf{b}^{h}-\mathbf{b}^{h-1} \Vert_2^2 + \lambda_5 \Vert \mathbf{c}^{h}-\mathbf{c}^{h-1} \Vert_2^2,
    \label{eq:lossBayesian}
  \end{aligned}
\end{equation}
where the $\ell_2$-norm is taken over the sum of $m$, i.e., $a^h = (a_1^h,...,a_M^h),\ \Vert a^{h}-a^{h-1} \Vert_2^2 = \sum_{m=1}^{M}(a_m^h-a_m^{h-1})^2$.
As before, the loss function (\ref{eq:lossBayesian}) can also be minimized by the \mbox{L-BFGS-B} algorithm \cite{LBFGSB1995} because the regularization terms are smooth.

\subsection{Multi-level sinogram interpolation algorithm}
\label{Sec: Algorithm}
\begin{algorithm}[htbp]
  \caption{Multi-level sinogram interpolation}
  \label{alg:multi_level_sino_interp}
  \begin{algorithmic}[1]
    \STATE \textbf{Input}: 
      \STATE$\quad$ 1. M: the number of sine waves to be recovered
      \STATE$\quad$ 2. Initial parameters $\mathbf{x}^{h_0} =\{\mathbf{A}^{h_0},\mathbf{\phi}^{h_0},\mathbf{a}^{h_0},\mathbf{b}^{h_0},\mathbf{c}^{h_0}\}$ 
    based on measurements of the secondary absorption sinogram at level $h_0=\lfloor\frac{H+1}{2}\rfloor$, 
      \STATE$\quad$ 3. $H+1$ levels of secondary intensity sinograms
      \STATE$\quad$ 4. level width $\zeta$ and $2\zeta$ levels of filtered secondary absorption sinograms
    \STATE \textbf{Output}: Optimized parameters $\mathbf{x}^{h} = \{\mathbf{A}^{h},\mathbf{\phi}^{h},\mathbf{a}^{h},\mathbf{b}^{h},\mathbf{c}^{h}\}$ at each level $h\in\{0,1,\dots,H\}$

    \STATE \textbf{Initialization}:
    \STATE $\quad$ Set initial parameters $\mathbf{x} \gets \mathbf{x}^{h_0}$

    \STATE \textbf{Step 1: Base level selection} 
    \FOR{$h\in(\lfloor\frac{H+1}{2}\rfloor-\zeta,\lfloor\frac{H+1}{2}\rfloor+\zeta)$}
      \STATE Compute parameters $\mathbf{x}^h$ for level $h$  secondary absorption sinogram by applying L-BFGS-B algorithm to (\ref{eq:lossfinal})
      \STATE If $\mathbf{x}^h$ satisfies $\mathbf{A}^{h}\neq 0$, $\mathbf{c}^{h}\neq 0$, and $\mathbf{a}^{h}\neq \mathbf{b}^{h}$ (if applicable, other requirements for the parameters), mark the $h$ as a feasible starting level. 
    \ENDFOR
    \STATE Select the floor of the median $h^{*}$ of all feasible starting levels

    \STATE \textbf{Step 2: Descending optimization}
    \STATE Initialize $\mathbf{x}^{h^{*}}$ and $\mathbf{d}^{h^{*}}$
    \FOR {$h = h^{*}-1$\ \textbf{to} $0$}
      \STATE Initialize $\mathbf{x}^{h}$ using $\mathbf{x}^{h+1}$, and $\mathbf{d}^{h}$ the corresponding gray value of the secondary background-only image
      \STATE Compute parameters $\mathbf{x}^h$ and $\mathbf{d}^{h}$ for level $h$ secondary intensity sinogram by applying L-BFGS-B algorithm to (\ref{eq:lossBayesian}), where $h$ and $h-1$ are switched
    \ENDFOR
    
    \STATE \textbf{Step 3: Ascending optimization}
    \STATE Initialize $\mathbf{x}^{h^{*}}$ and $\mathbf{d}^{h^{*}}$
    \FOR {$h = h^{*}+1$\ \textbf{to} $H$}
      \STATE Initialize $\mathbf{x}^{h}$ using $\mathbf{x}^{h-1}$, and $\mathbf{d}^{h}$ the corresponding gray value of the secondary background-only image
      \STATE Compute parameters $\mathbf{x}^h$ and $\mathbf{d}^{h}$ for level $h$ secondary intensity sinograms by applying L-BFGS-B algorithm to (\ref{eq:lossBayesian})
    \ENDFOR

    \STATE \textbf{Step 4: Reconstruction and interpolation}
    \STATE Integrate the parameters into model (\ref{eq:discreteBG})
    \STATE Compute interpolated sinograms at arbitrary angles
    \RETURN $\{\mathbf{x}^{h},\mathbf{d}^{h}\}_{h=0}^H$
  \end{algorithmic}
\end{algorithm}

While the choice of starting levels remains flexible—allowing for the selection of slices with the most distinct sinusoidal structures to ensure accurate parameter estimation—our observations of the secondary sinograms along the $h$-axis indicate that levels near the center, i.e., $\lfloor \frac{H+1}{2} \rfloor$, exhibit the clearest sine waves. Consequently, we employ these central sinograms as candidates for the starting point. 


Among these candidates, we aim for a starting level $h^{*}$ for which one clearly observes at least one sine wave, that is,  the amplitude parameters $\mathbf{A}$ and the gray values $\mathbf{c}$ do not all vanish, and the intervals $[a_m,b_m)$ do not all collapse to a point. To check for feasibility, our base level selection step applies the L-BFGS-B algorithm to (\ref{eq:lossfinal}) without additional regularization for all candidate levels, where, for the initialization, we use measurements obtained for the center level, e.g., based on visual inspection of the fully sampled primary sinograms and the subsampled secondary sinograms. Among the identified feasible levels  $h$, we choose the floor of the median $h^{*}$. Alternatively, one could also employ stronger requirements, e.g., ask for a larger number of non-vanishing sine waves. 
This yields the final interpolation procedure given in Algorithm \ref{alg:multi_level_sino_interp}.

While for our data, a single starting level suffices, it may sometimes be beneficial to use multiple starting levels and use each of them for a reconstruction procedure in its $h$-neighborhood. Furthermore, developing an automated process for this selection presents an intriguing direction for future research.

Note that the procedure described above is not guaranteed to yield a feasible starting level $h^{*}$, especially when the initial parameters are distant from the optimal solution.
Potential remedies include increasing the width $\zeta$ or tuning the initialization for the base level selection.

\section{Numerical results}
\label{chapter:Numerical-Experiments}
In this section, we will test  Algorithm \ref{alg:multi_level_sino_interp} on a real dataset consisting of primary and secondary scans of a $1.6\times 0.8\times 0.8$ mm sized coil acquired by a micro-CT system. 


As the first and last secondary scans in the dataset are imaged at the same angle, they are expected to be the same. Hence, we will not process the secondary scan at the angle $2\pi$  in our sinogram reconstruction.



\subsection{Recovery with multi-level sinogram interpolation algorithm}
Our processed dataset contains 4 folders:
\begin{enumerate}
  \item Primary scans: $J+1=1001$ equidistantly sampled primary scans in $[0,2\pi]$. Every scan was imaged 5-20 times, then averaged for denoising.
  \item Original secondary scans for assessing the quality of the reconstruction (``oracle information''): $J+1=1001$ equidistantly sampled secondary scans in $[0,2\pi]$. Every scan was imaged 1-3 times, then averaged for denoising. 
  \item Subsampled secondary scans to be processed by the reconstruction algorithm: $N+1$ equidistantly sampled secondary scans in $[0,2\pi]$. 
  \item  Background-only images: primary and secondary scans without a sample (only one angle required, as there is no object to be rotated).
\end{enumerate}
For every secondary scan, the mask is applied directly via multiplication. 
If folder 2 is given to compute the interpolation error, folder 3 could be subsampled from 2 if $N\mid J$.
Since the secondary sinograms are very noisy and has striped artifact after subtracting the constant background (see Figure \ref{MinusSecSino}),
we apply a low-pass filter with a cutoff frequency of 0.04 for each column of the secondary sinograms.
The scaling factor of the Cauchy kernel is set to $\sigma = 9$. The regularization parameters are set to $[\lambda_1,\lambda_2,\lambda_3,\lambda_4,\lambda_5]=[100,1,1,1,1]$ in base level selection, and $[\lambda_1,\lambda_2,\lambda_3,\lambda_4,\lambda_5]=[100,10,1,1,1]$ in descending and ascending optimization.
We choose $M = 2$ due to two prominent sinusoids and parameter initialization as follows

\begin{table}
  \centering
  \begin{tabular}{ |c|c|c| } 
  \hline
  Parameters & Initialization & Recovered (h=172) \\
  \hline
  $\mathbf{A}$ & [30, 30] & [25.548, 31.359]\\
  \hline
  $\mathbf{\phi}$ &  [0, 3.141] & [0, 3.289]\\ 
  \hline
  $\mathbf{a}$ &   [140, 140] & [141.677, 140.249]\\ 
  \hline
 $\mathbf{b}$ &   [180, 180] & [185.755, 185.635]\\ 
  \hline
  $\mathbf{c}$ & [1.5, 1.5] & [1.977, 2.142]\\
  \hline
  \end{tabular}
  \caption{Initial and recovered parameters of model (\ref{eq:finaldiscrete}) for secondary sinograms recovery}
\end{table}

The sinogram interpolation algorithm \ref{alg:multi_level_sino_interp} is tested with $N\in \{2,4,5,8,10,25\}$ 
subsampled secondary projections and level width $\zeta = 10$. 

\begin{figure}
  \centering
  \includegraphics[scale=0.65]{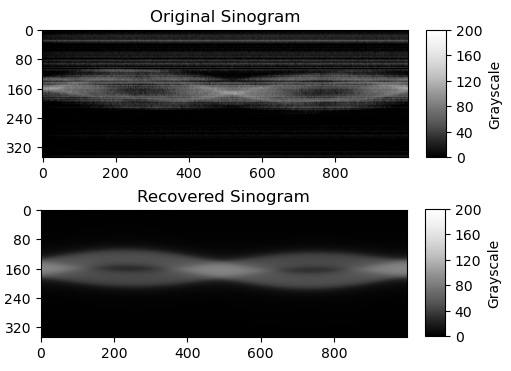}
  \caption{Level $h = 172, H = 348$, the successfully reconstructed secondary absorption sinogram during the base level selection step in Algorithm \ref{alg:multi_level_sino_interp} ($N = 25$).}
  \label{Recon172}
\end{figure}
The original absorption sinogram in Figure \ref{Recon172} 
is based on original secondary scans 
with the number of angles $J= 1000$, but only subsampled and filtered secondary absorption sinograms at $N = 25$ equidistant angles are used for optimization.

\begin{figure}
  \centering
  \includegraphics[scale=0.65]{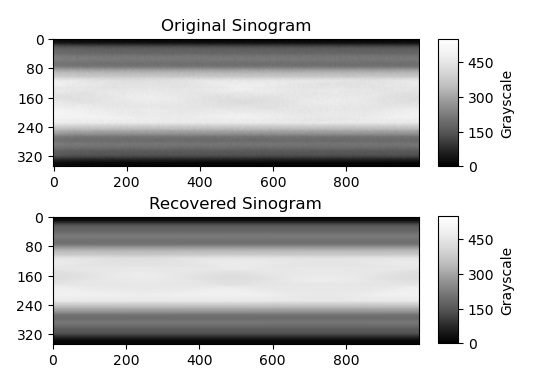}
  \caption{level $h = 159, H = 348$, the successfully reconstructed secondary intensity sinogram during the descending optimization step in Algorithm \ref{alg:multi_level_sino_interp} ($N = 25$).}
  \label{Recon160}
\end{figure}

\begin{figure}
  \centering
  \includegraphics[scale=0.65]{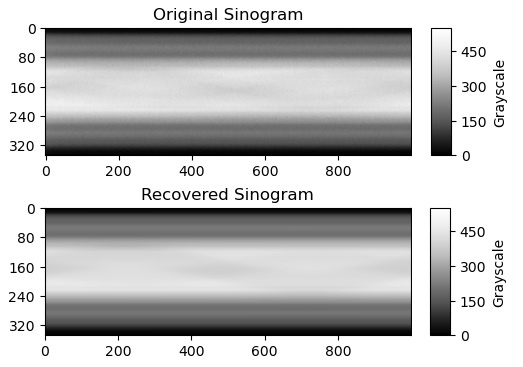}
  \caption{level $h = 199, H = 348$, the successfully reconstructed secondary intensity sinogram during the ascending optimization step in Algorithm \ref{alg:multi_level_sino_interp} ($N = 25$).}
  \label{Recon199}
\end{figure}

\begin{figure}
    \centering

    \begin{minipage}[t]{0.48\linewidth}
        \centering
        \includegraphics[width=\linewidth]
        {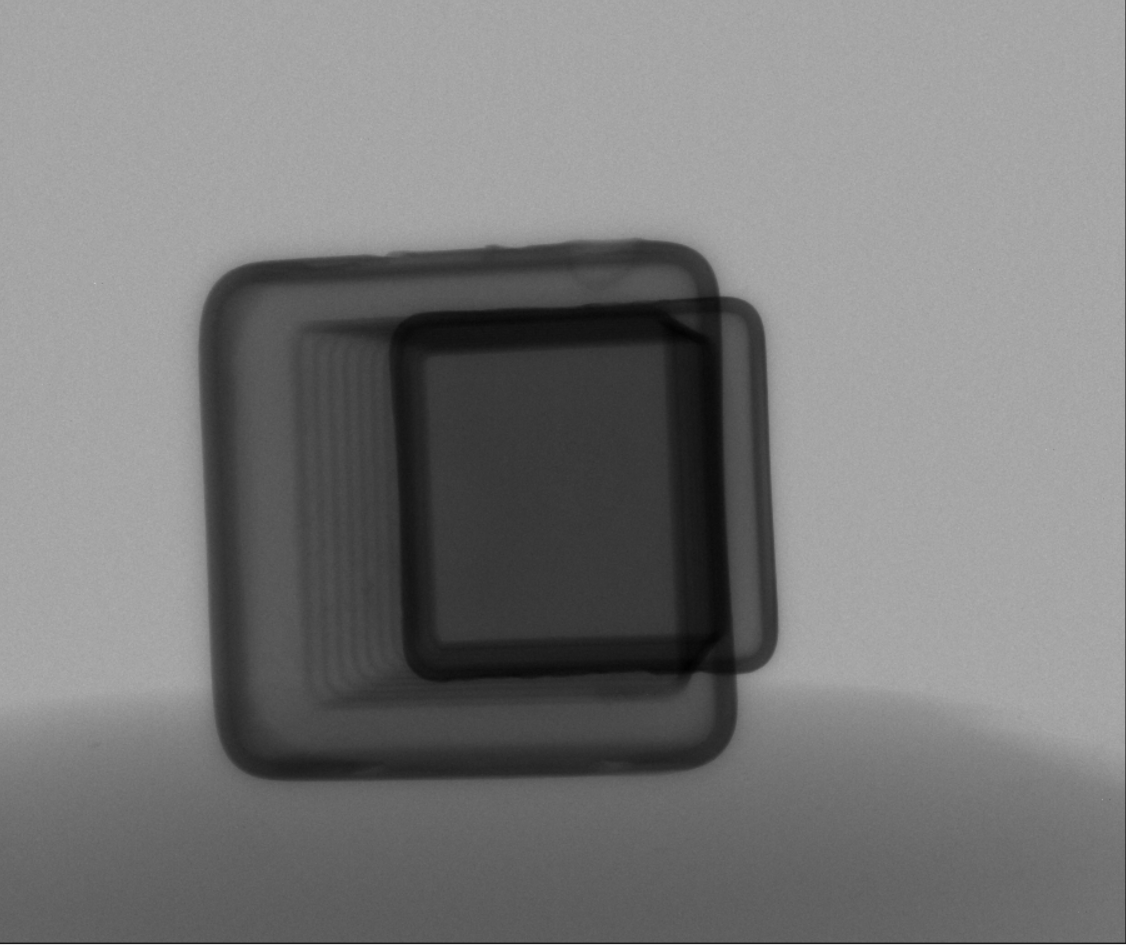}

        \smallskip
        {\small (a) Corrected with direct secondary scan information}
    \end{minipage}
    \hfill
    \begin{minipage}[t]{0.48\linewidth}
        \centering
        \includegraphics[width=\linewidth]
        {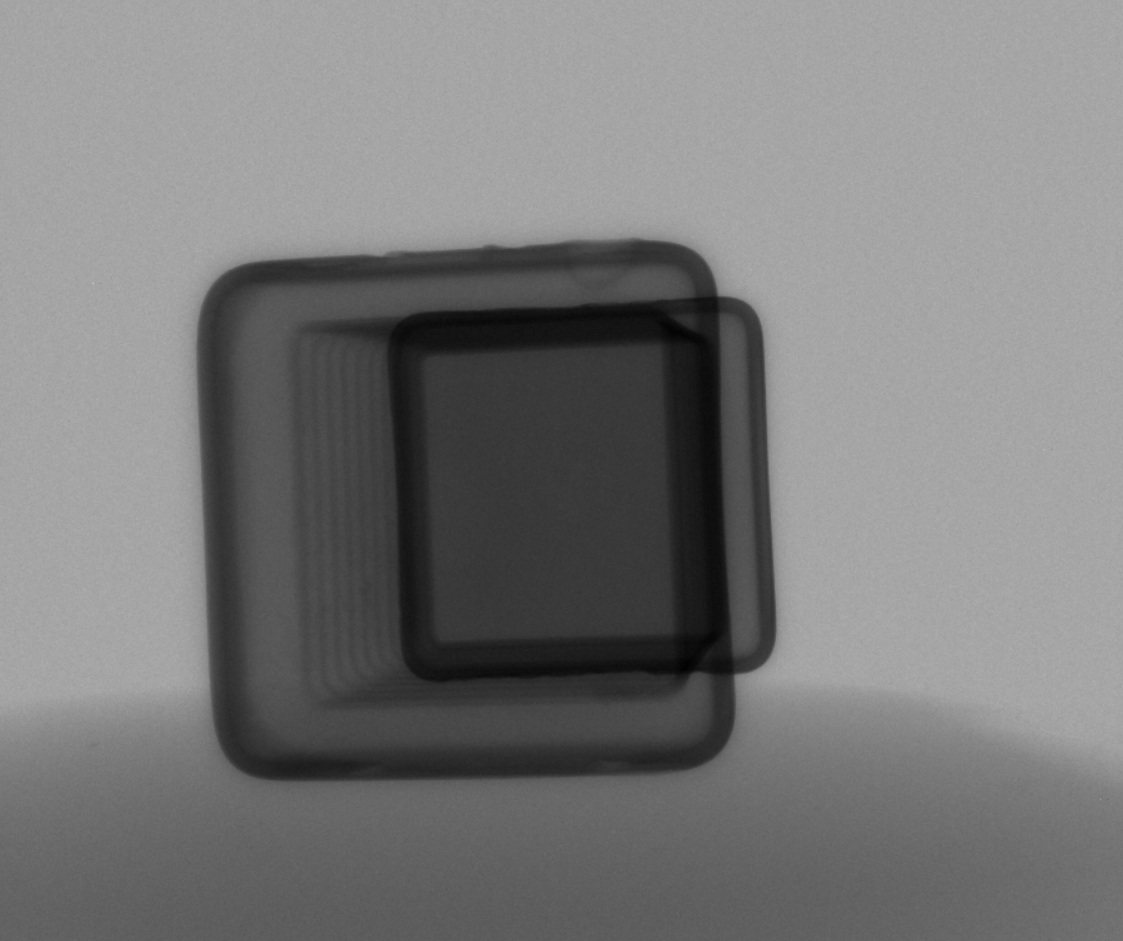}

        \smallskip
        {\small (b) Corrected with interpolated secondary scan information}
    \end{minipage}

    \caption{Corrected images at $j=43$, number of total angles $J=1000$,
    number of subsampled angles $N=25$.}
    \label{fig:corrected}
\end{figure}

After getting a set of corrected images, FDK reconstruction algorithm is used to obtain the 3D volumetric information of the object.

\begin{figure}[htbp]
    \centering

    \begin{minipage}[t]{0.47\linewidth}
        \centering
        \includegraphics[width=\linewidth]{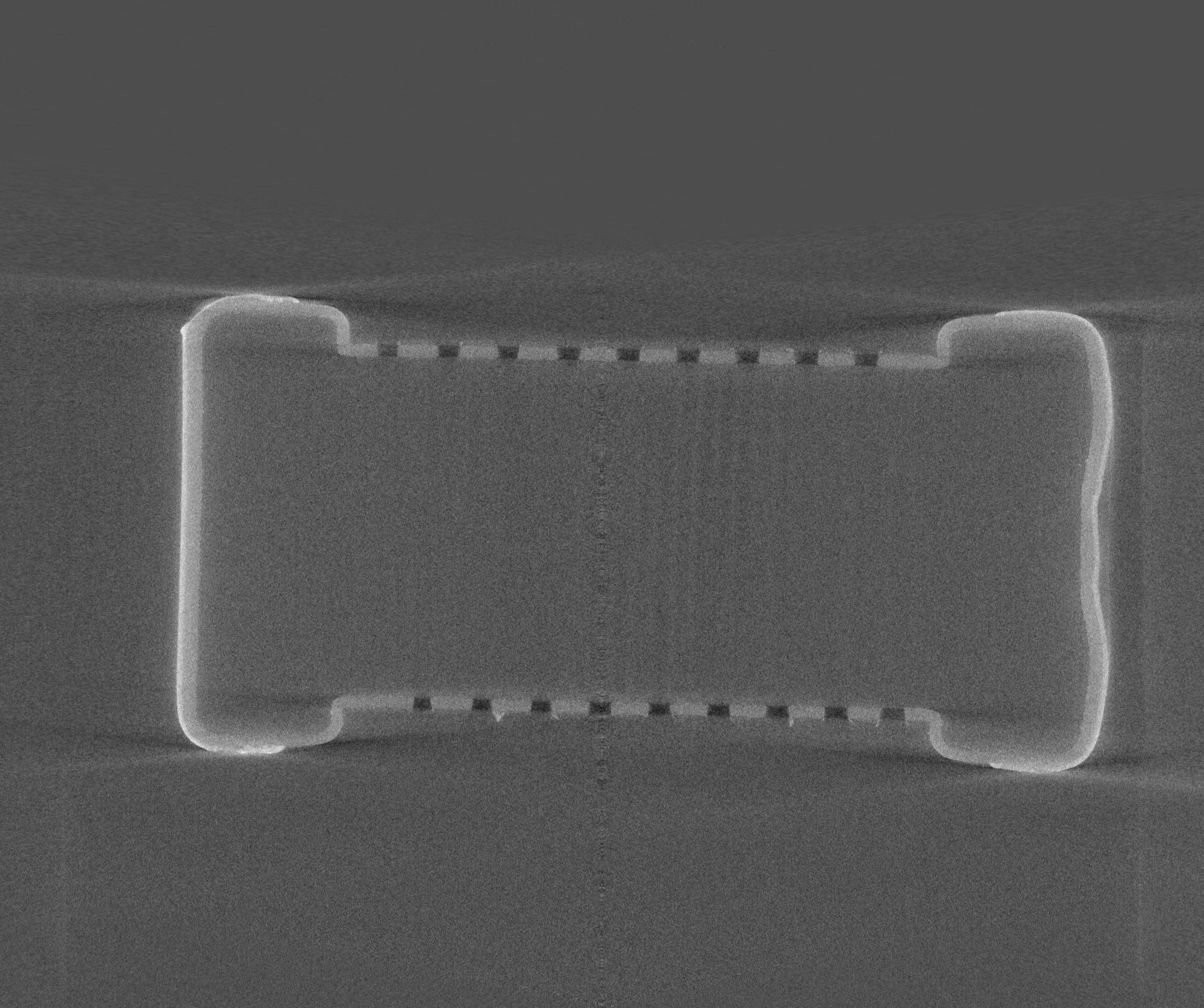}

        \smallskip
        {\small (a) Projection along the x-axis}
    \end{minipage}
    \hfill
    \begin{minipage}[t]{0.47\linewidth}
        \centering
        \includegraphics[width=\linewidth]{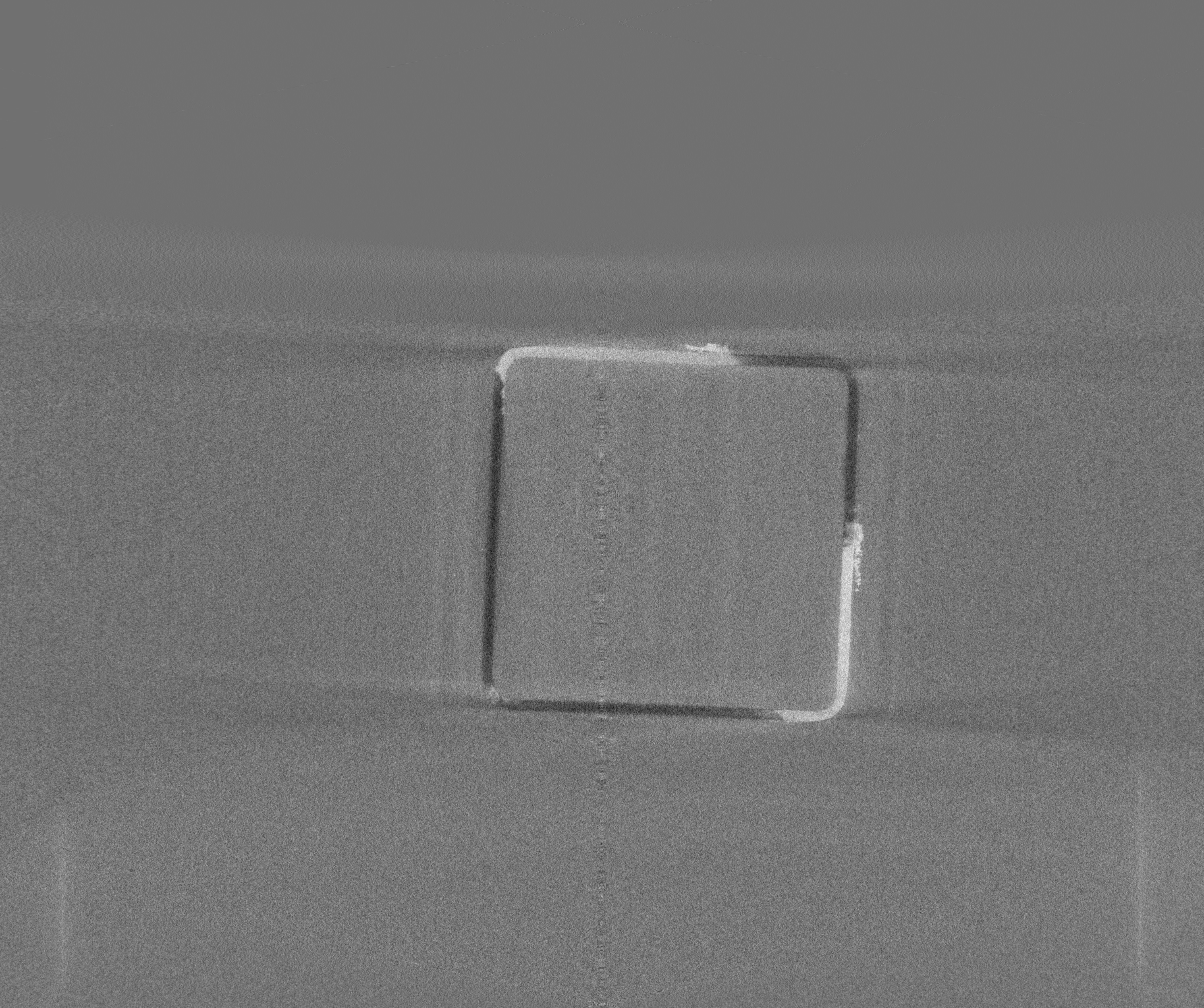}

        \smallskip
        {\small (b) Projection along the y-axis}
    \end{minipage}

    \par\medskip

    \begin{minipage}[t]{0.47\linewidth}
        \centering
        \includegraphics[width=\linewidth]{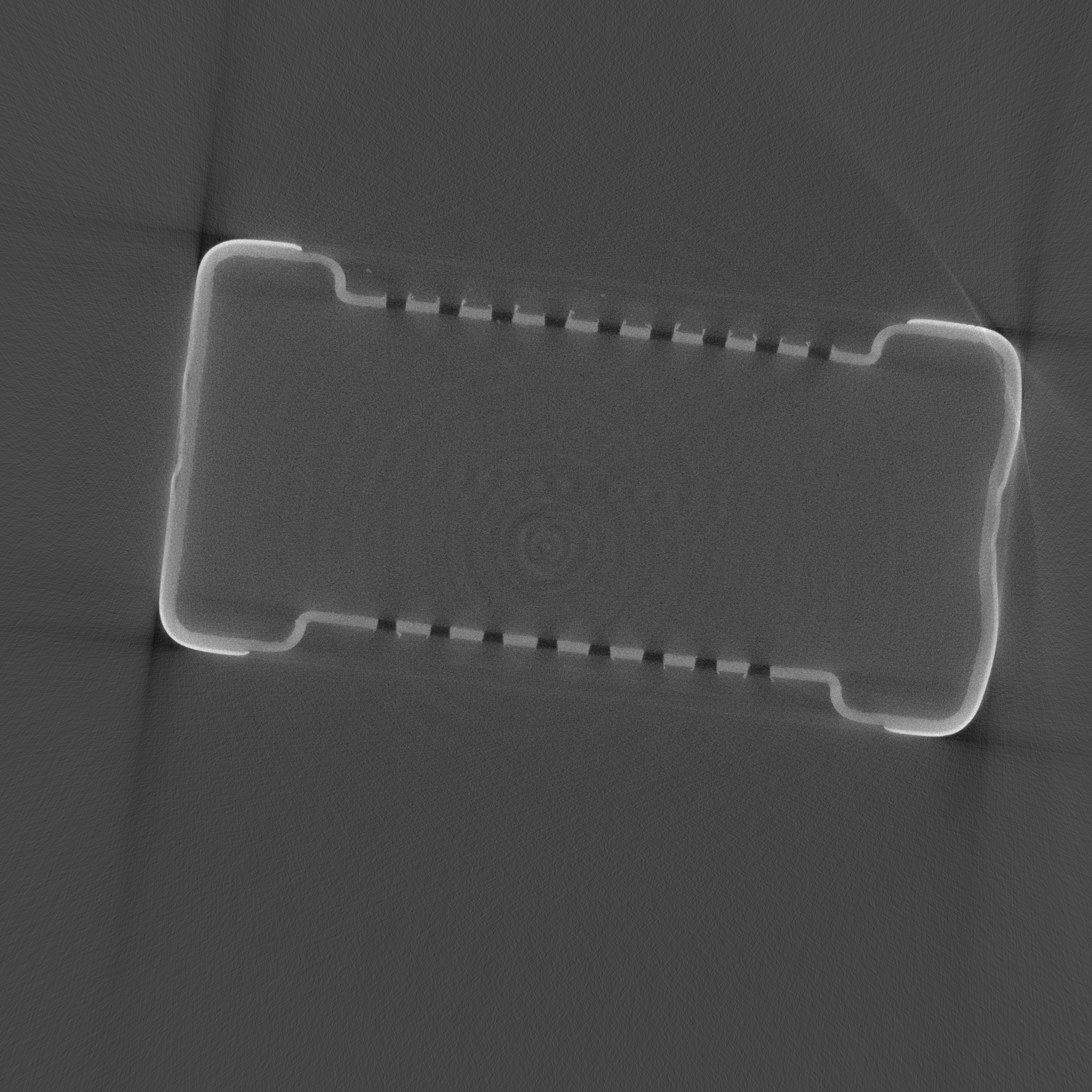}

        \smallskip
        {\small (c) Projection along the z-axis}
    \end{minipage}

    \caption{Reconstructed object using the FDK algorithm. Corrected images
    are processed with interpolated secondary scans
    ($N=25$, $J=1000$, slice $1261$ of $2502$).}
    \label{fig:3_subfigures}
\end{figure}
\begin{figure}[htbp]
    \centering

    \includegraphics[width=0.85\linewidth]
    {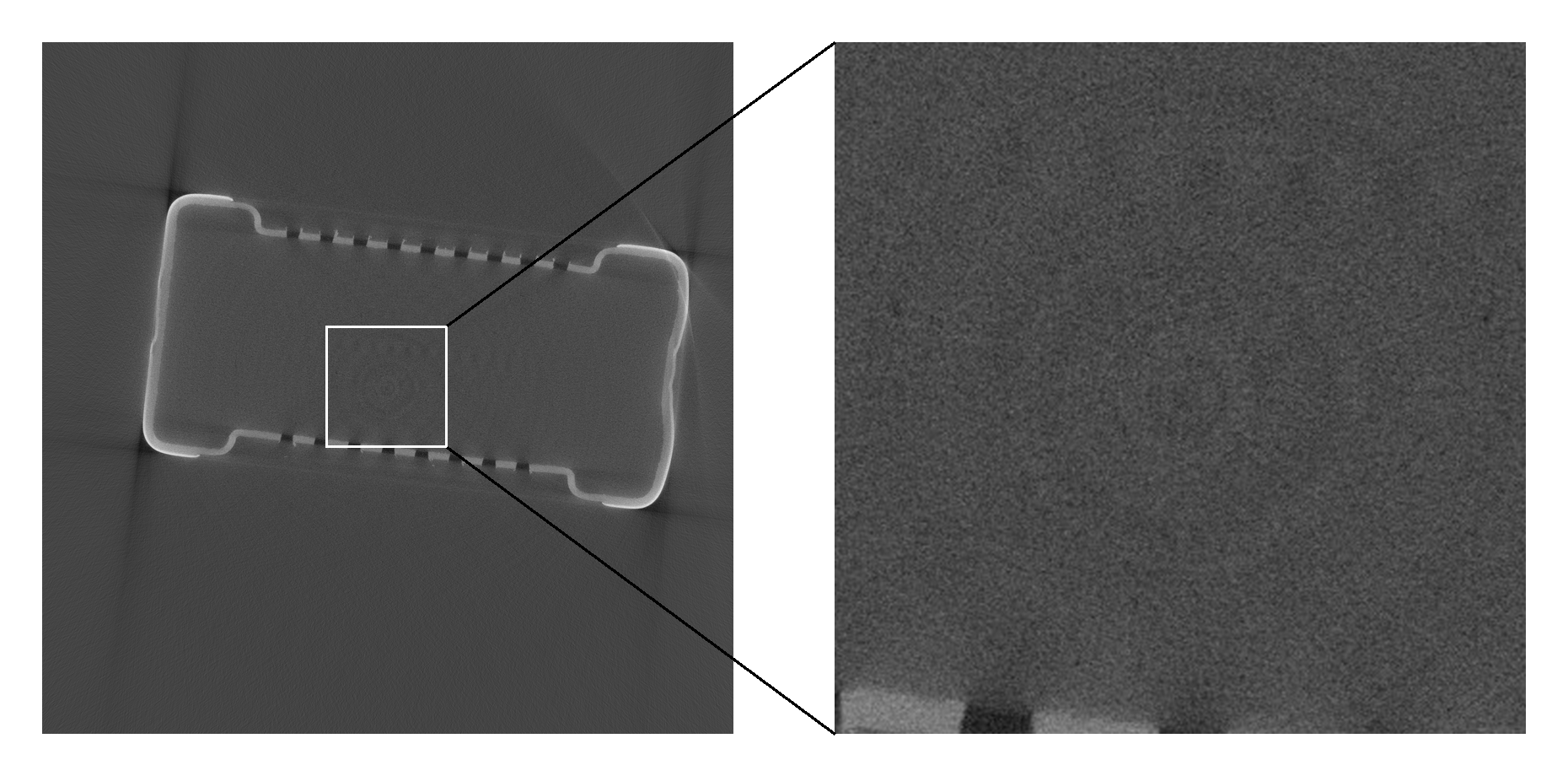}

    \smallskip
    {\small (a) Using acquired secondary scans}

    \par\medskip

    \includegraphics[width=0.85\linewidth]
    {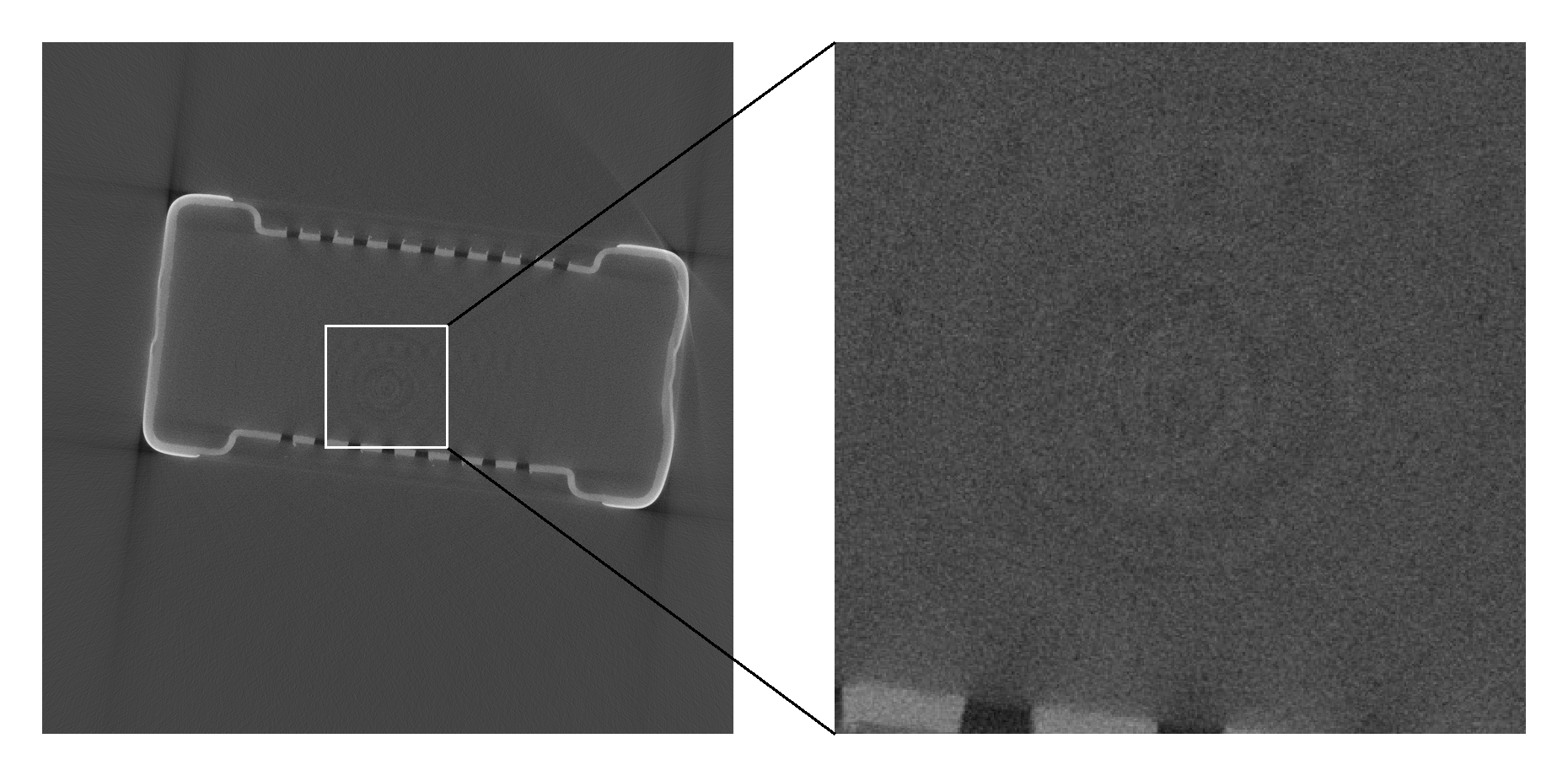}

    \smallskip
    {\small (b) Using interpolated secondary scans ($N=25$)}

    \par\medskip

    \includegraphics[width=0.85\linewidth]
    {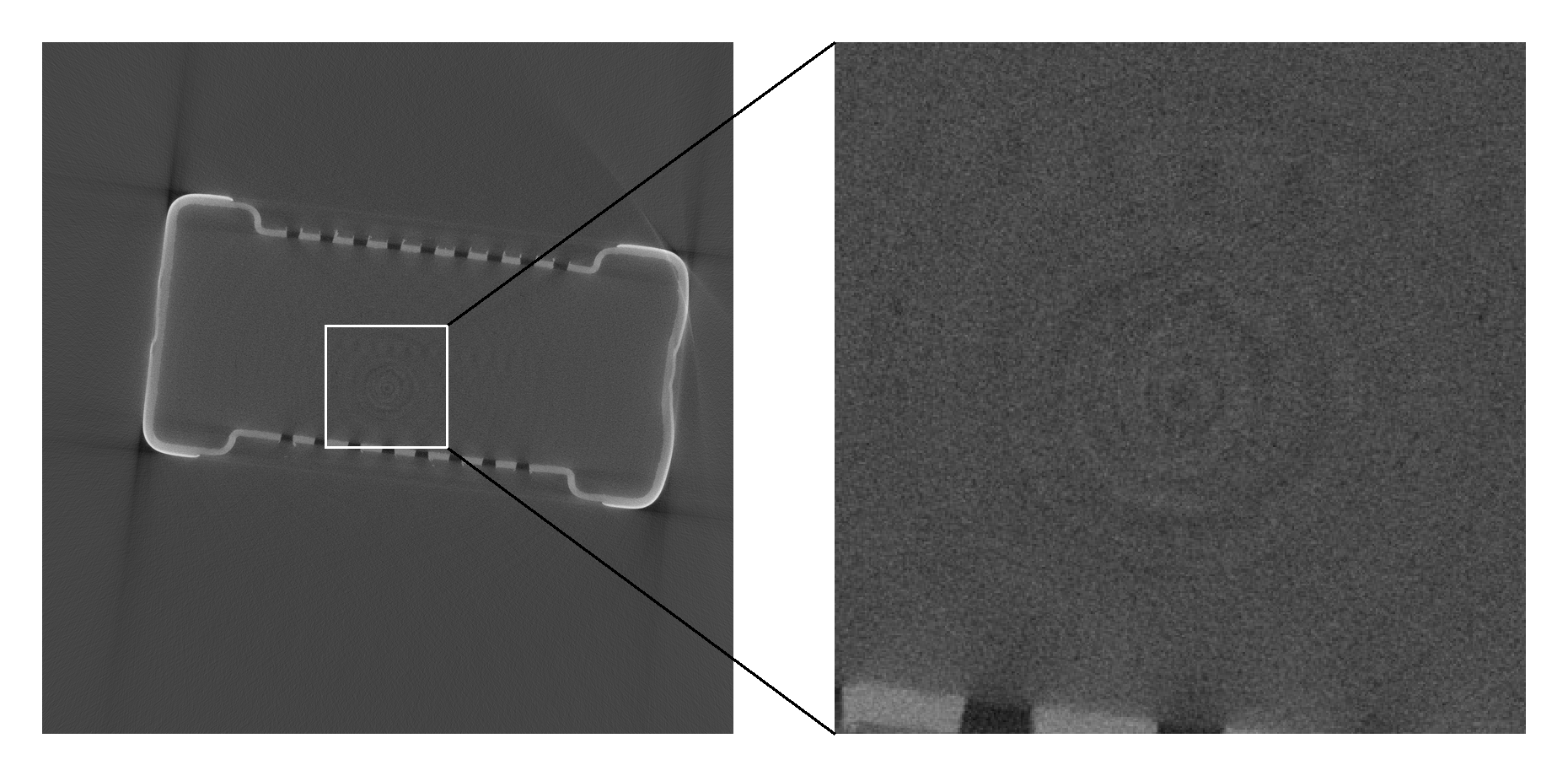}

    \smallskip
    {\small (c) Using interpolated secondary scans ($N=5$)}

    \par\medskip

    \includegraphics[width=0.85\linewidth]
    {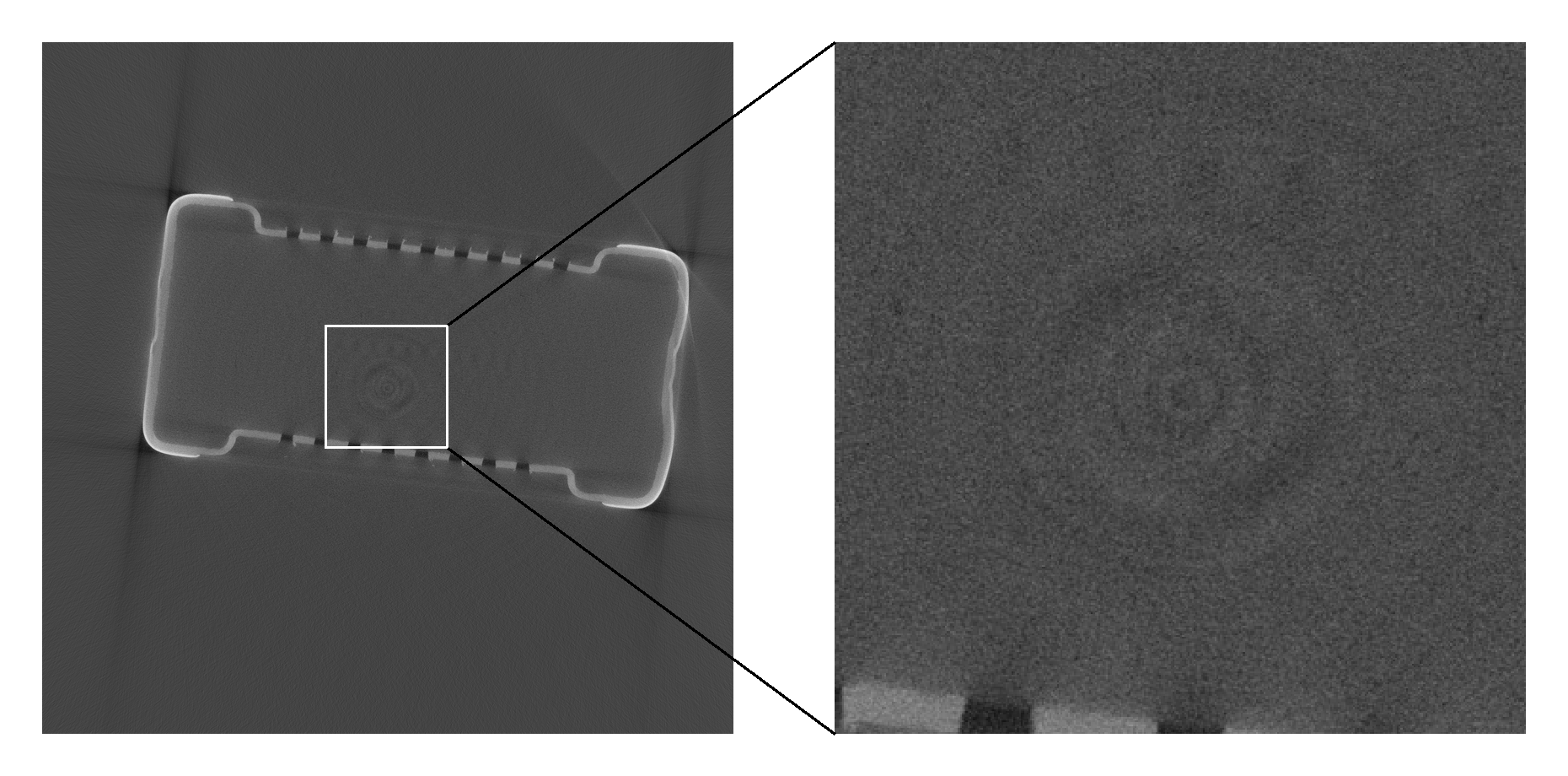}

    \smallskip
    {\small (d) Using interpolated secondary scans ($N=4$)}

    \par\medskip

    \includegraphics[width=0.85\linewidth]
    {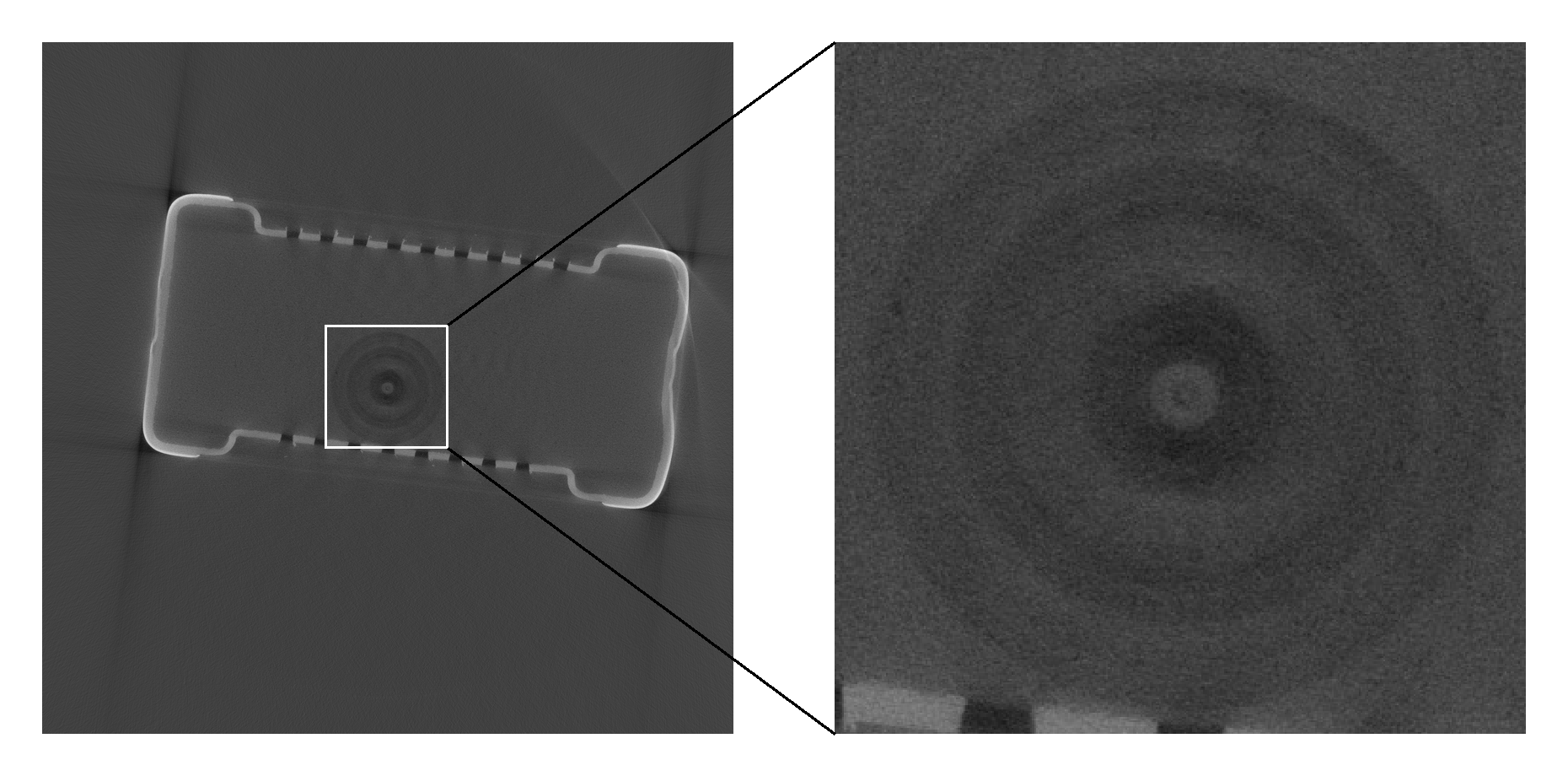}

    \smallskip
    {\small (e) Reconstructed using only primary scans}

    \caption{Reconstructed object along the $z$-axis using the FDK algorithm
    (slice $1261$ of $2502$) for different secondary-scan configurations.
    The enlarged regions show the same area of interest for all reconstructions.}
    \label{recon_compare}
\end{figure}

Figures~\ref{fig:corrected} and \ref{fig:3_subfigures} show that the secondary radiation artifact is eliminated using both the original secondary scans and the interpolated ones with $N = 25$.
Figure~\ref{recon_compare} shows slices of the reconstructed $3D$ volume along the z-axis for different values of N. For the dataset considered in this study, a noticeable degradation in reconstruction quality is observed when N is reduced from $5$ to $4$. This behavior is specific to the present dataset, and the minimum number of secondary scans required to preserve reconstruction quality may vary depending on the object and the characteristics of the measured data.

\section{Conclusions}
In this paper, we introduced a novel approach for modeling and recovery of secondary sinograms in micro-CT imaging.
Our main contributions are a novel continuous model for the secondary sinograms and a corresponding interpolation algorithm. Our approach allows to easily reconstruct  
secondary sinograms from sparsely sampled angles using simple box constraints and classical optimization methods with only a few parameters.

For future work, we find it an interesting problem to fully automate the parameter initialization and to develop our method for ill-posed sinogram interpolation problems without any information about the object attenuation.
One could also explore anisotropic objects with more complicated sine waves whose gray values and width vary with the angular parameter $\theta$. 

\section{Acknowledgments}
The authors gratefully acknowledge the generous support and resources provided by Waygate Technologies, e.g., Baker Hughes Digital Solutions GmbH. This research was made possible by the economic support provided during the project, as well as crucial access to real-world datasets and the reconstruction software Datos$|$x utilized for model development and analysis. The authors particularly thank the organization for facilitating the professional industrial context necessary for this work. 
Furthermore, FK and AL acknowledge support by the German Federal Ministry of Research, Technology, and Space and the State of Bavaria in the context of the Munich Center for Machine Learning.

\bibliographystyle{IEEEtran}
\bibliography{literature}

\end{document}